\documentclass{aa}
\usepackage{epsfig}
\nonstopmode
\newcommand{\Egamma}{E_\gamma}
\newcommand{\Eline}{E_{\rm line}}

\def\finaldraft#1{{#1}}

\begin{document}

\title{Radiation Transfer of Emission Lines in Curved Space-Time} 

\authorrunning{Fuerst and Wu}
\titlerunning{Radiation Transfer of Emission Lines in Curved Space-Time}

   \author{ Steven V.~Fuerst  and Kinwah Wu 
}

   \offprints{S.~V.~Fuerst}

   \institute{Mullard Space Science Laboratory, University College London,
              Holmbury St Mary, Surrey RH5 6NT, UK \\
              \email{svf@mssl.ucl.ac.uk, kw@mssl.ucl.ac.uk}       
             }

\date{Received:    }

\abstract{
  We present a general formulation for ray-tracing calculations 
    in curved space-time. 
  The formulation takes full account of relativistic effects  
    in the photon transport   
    and the relative motions of the emitters 
    and the light-of-sight absorbing material.  
  We apply the formulation to calculate the emission   
    from accretion disks and tori around rotating black holes.   
  In our model the emission lines and continuum 
    originate from an accretion disk or torus,  
    and the motion of the emitters in the disk/torus is  
    determined by the gravity of the black hole  
    and the space-time structure near the black hole. 
  The line-of-sight absorbing medium is comprised 
    of cold absorbing cloudlets.
  These cloudlets are kinematically hot, 
    with their velocity dispersion 
    determined by the local virial temperature. 
  The emission from the accretion disk/torus is resonantly absorbed/scattered.  
  Our calculations demonstrate that 
    line-of-sight absorption significantly modifies 
    the profiles of lines from the accretion disks.  
  It is often difficult to disentangle absorption effects  
    from other geometrical and kinematics effects, 
    such as the viewing inclination and the spin of the black hole.  
  Our calculations also show that 
    emission lines from accretion tori and from thin accretion disks 
    differ substantially. 
  Large geometrical obscuration could occur in tori, 
    and as a consequence 
    lines from tori generally have much weaker redshift wings 
    at large viewing inclination angles. 
   Moreover, the blue peak is truncated. 
\keywords{
  accretion, accretion disks ---
  black hole physics ---
  galaxies: active ---
  line: profiles ---
  radiative transfer ---
  relativity  }
}

   \maketitle
%

\section{Introduction}

The strong X-rays observed in active galactic nuclei (AGN) 
  and some X-ray binaries are believed to be powered 
  by accretion of material into black holes. 
The curved space-time around the black hole 
  influences not only the accretion hydrodynamics     
  but also the transport of radiation from the accretion flow. 
  
Emission lines from thin Keplerian disks around non-relativistic stellar objects 
  generally have two symmetric peaks (Smak 1969),  
  corresponding to the approaching and receding line-of-sight velocities 
  due to disk rotation.  
Because of various relativistic effects,  
  lines from accretion disks around black holes
  do not always have symmetrical double-peak profiles.   
The accretion flow near a black hole is often close to the speed of light, 
  and emission is relativistically boosted. 
The blue peak of the line therefore becomes stronger and sharper.  
Moreover, the strong gravity near the black-hole event horizon causes time dilation, 
  which shifts the line to lower energies.  
Emission lines from accretion disks around black holes 
  appear to be broad, with a very extended red wing and a narrow, sharp blue peak   
  (see e.g. the review by Fabian et al.\ (2000) and references therein).     
Furthermore, gravitational lensing can produce multiple images 
  and self-occultation, further modifying the emission line profile.  

Various methods have been used to calculate the profiles of emission lines 
  from accretion disks around black holes.  
The methods can be roughly divided into three categories. 
We now discuss each of them briefly. 
The first method uses a transfer function 
  to map the image of the accretion disk onto a sky plane 
  (Cunningham 1975, 1976).
The accretion disk is assumed to reside in the equatorial plane. 
It is Keplerian and geometrically thin, but optically thick.    
The space-time metric around the black hole is first specified, 
  and the energy shift of the emission (photons) 
  from each point on the disk surface is then calculated.  
A parametric emissivity law for the disk emission is usually used 
  --- typically, a simple power-law which decreases radially outward. 
The specific intensity at each point in the sky plane 
  is determined from the energy shift  
  and the corresponding specific intensity at the disk surface,  
  using the Lorentz-invariant property.  
The transfer-function formulation (Cunningham 1975, 1976) 
  has been applied to line calculations in settings 
  ranging from thin accretion rings (e.g.\ Gerbal \& Pelat 1981) 
  and accretion disks around Schwarzschild (e.g.\ Laor 1991) 
  and rotating (Kerr) black holes (e.g.\ Bromley, Chen \& Miller 1997).  
The second method makes use of the impact parameter of photon orbits 
  around Schwarzschild black holes 
  (e.g.\ Fabian et al. 1989; Stella 1990; Kojima 1991).  
The transfer function in this method 
  is described in terms of elliptical functions,  
  which are derived semi-analytically.
The Jacobian of the transformation from the accretion disk to sky plane 
  is, however, determined numerically 
  via infinitesimal variations of the impact parameter (Bao 1992).   
The method can be generalized to the case of rotating black holes 
  by using additional constants of motion 
  (Viergutz 1993; Bao, Hadrava \& Ostgaard 1994; 
  Fanton et al.\ 1997; Cadez, Fanton \& Calvani 1998).
The third method simply considers direct integration of the geodesics 
  to determine the photon trajectories and energy shifts    
  (Dabrowski et al.\ 1997; Pariev \& Bromley 1998; Reynolds et al.\ 1999). 

These calculations have shown how  
  the dynamics of the accretion flow around the black hole 
  and the curved space-time shape the line profiles.   
Various other aspects of the radiation processes,  
  e.g.\ reverberation and reflection (Reynolds et al.\ 1999) and  
  disk warping (Cadez et al.\ 2003)  
  were also investigated using the methods described above.   
The results obtained from these calculations 
  have provided us with a basic framework 
  for interpreting X-ray spectroscopic observations, 
  in particular, the peculiar broad Fe K$\alpha$ lines  
  in the spectra of AGN, e.g.\ MCG-6-30-15 (Tanaka et al.\ 1995).   
While existing studies have put emphasis on the energy shift of the emission, 
  transport effects such as extinction have been neglected.     
Resonant absorption (scattering) by ambient material 
  can greatly modify the disk emission line profile. 
This effect was already demonstrated in a study by Ruszkowski \& Fabian (2000), 
  in which a simple rotating disk-corona provides the resonant scattering.   
 
Here, we present ray-tracing calculations of spectra
  from relativistic flows in curved space-time.
We include line-of-sight extinction and emission explicitly 
  in the formulation.   
The radiative-transfer equation is derived 
  from the Lorentz-invariant form of the conservation law.  
It reduces to the standard classical radiative-transfer equation 
  in the non-relativistic limit.    
The formulation can incorporate dynamical and geometric models 
  for the line-of-sight absorbing and emitting material. 
As an illustration, we calculate the \finaldraft{emision line profiles} 
  from thin accretion disks and thick accretion tori 
  around rotating black holes.
The emitted spectra include a power law continuum together
  with a line.
This emission is resonantly scattered by the line-of-sight-material.
We include the contribution from higher-order images and allow for self-occultation.

We organize the paper as follows. 
In \S2 we show the derivation of the transfer equation.  
In \S3 we construct the equation of motion for
  free particles in a Kerr space-time   
  and for force-constrained particles 
  for some simple parametric models. 
In \S4 we construct a thin disk and a thick torus model.       
In \S5 we generalize this by adding in absorption due to a distribution of absorbing clouds.
In \S6 we present the results from the models where either
  emission geometry (tori), or absorption (clouds) are important.

\section{Radiative-Transfer Equation} 

Throughout this paper, we adopt the usual convention $c=G=h=1$  
  for the speed of light, gravitational constant and Planck constant. 
The interval in space-time is specified by 
\begin{equation}
\label{metric} 
 d\tau^2 = g_{\alpha \beta} dx^{\alpha} dx^{\beta} \   
\end{equation}  
  where $g_{\alpha \beta}$ is the metric.   

Consider a bundle of particles which fill a phase-space volume element   
\begin{equation}
\label{bundle}
  {d\cal{V}} \equiv dx\,dy\,dz\,dp^x\,dp^y\,dp^z\ ,
\end{equation}
  where $dx\,dy\,dz (\equiv dV)$ is the three-volume 
  and $dp^x\ dp^y\ dp^z$ is the momentum range, at a given time $t$.   
Liouville's Theorem reads 
\begin{equation}
   \frac{d{\cal V}}{d\lambda} = 0      
\end{equation} 
  (see Misner, Thorne \& Wheeler 1973), 
  with $\lambda$ here being the affine parameter 
  for the central ray in the bundle.   
The volume element $d{\cal V}$ is thus Lorentz invariant. 
 
The distribution function for the particles in the bundle, $F(x^i, p^i)$ 
  is given by
\begin{equation}
  F(x^i, p^i) = {dN \over d{\cal V}}\ ,
\end{equation}
  where $dN$ is the number of particles in the three-volume.  
Since $dN/d{\cal V}$ is Lorentz invariant, 
   $F(x^i, p^i)$ is Lorentz invariant.
From equation (\ref{bundle}), we have
\begin{equation}
\label{rawfluxdefn}
  F={dN\over p^2 dV\,dp\,d\Omega}\ ,
\end{equation}
  where $p^2\,dp\,d\Omega=dp^x\,dp^y\,dp^z$. 
For massless particles, $v = c = 1$ and $\vert p\vert=E$.  
The number of photons in the given spatial volume 
  is therefore the number of photons flowing through an area $dA$ in a time $dt$.  
It follows that 
\begin{equation}
\label{flux_inv}
  F={dN\over E^2dA\,dt\,dE\,d\Omega}\ .
\end{equation}
Recall that the specific intensity of the photons is  
\begin{equation}
\label{inten_inv}
   I_\nu={E dN\over dA\,dt\,dE\,d\Omega}\ ,
\end{equation}
By inspection of equations (\ref{flux_inv}) and (\ref{inten_inv}), we obtain
\begin{equation}
  F=\frac{I_\nu}{E^3}=\frac{I_\nu}{\nu^3}\ ,  
\end{equation} 
  where $\nu ~(= E)$ is the frequency of the photon.  
We will use this Lorentz invariant intensity, ${\cal I}\equiv F$, 
  in the radiative transfer formulation.

In a linear medium, extinction is proportional to the intensity, 
  and the emission is independent of the intensity of the incoming radiation.  
The radiative transfer equation is therefore    
\begin{equation}
\label{classradtrans}
  \frac{d{\cal I}}{d s}=-\chi{\cal I} + \eta\left(\frac{\nu_0}{\nu}\right)^3 \ ,  
\end{equation}
   where $\chi$ is the absorption coefficient, 
  $\eta$ is the emission coefficient 
  and $ds$ is the length element the ray traverses.    
The equation in this form is defined in the observer's frame, 
  and the absorption and emission coefficients 
  are related to their counterparts in the rest frame with respect to the medium via   
\begin{eqnarray}
\label{chframe}
  \chi&=&\left(\frac{\nu_0}{\nu}\right)\chi_0  \ , \\
  \eta&=&\left(\frac{\nu}{\nu_0}\right)^2\eta_0  \ ,  
\end{eqnarray}  
  where the subscript ``0'' denotes quantities in the local rest frame.  

The relative energy / frequency shift in a moving medium with respect to an observer at infinity is given by 
\begin{equation}
\label{freqshift} 
   \frac{E_0}{E}=
   \frac{\nu_0}{\nu}=
   \frac{p^\alpha u_\alpha\vert_\lambda}{p^\alpha u_\alpha\vert_\infty}\ , 
\end{equation}
  where $u^\alpha$ is the four-velocity of the medium 
  as measured by an observer, and  
\begin{eqnarray}
\label{chframe2}
  \frac{ds}{d\lambda}&=&-p^\alpha u_\alpha\vert_\infty\ . 
\end{eqnarray}  
The radiative transfer equation (equation [\ref{classradtrans}]) 
 in the co-moving frame is therefore 
\begin{equation}
\label{radtrans}
\frac{d{\cal I}}{d\lambda}
  =-p^\alpha u_\alpha\vert_\lambda 
   \left[-\chi_0(x^\beta,\nu){\cal I}+\eta_0(x^\beta, \nu)\right]\  
\end{equation} 
  (see Baschek et al.\ 1997).  

The results in the co-moving frame can be used 
  to determine the intensity and frequency in the other reference frames.   
The ray is specified by choosing $x^\alpha(\lambda_0)$ and $p^\alpha(\lambda_0)$.  
From the geodesic equation, 
  we have $d p^\alpha/d \lambda+\Gamma^\alpha_{\beta\gamma}p^\beta p^\gamma=0$, where we have scaled $\lambda$ by $m$ for massive,
  and by $1$ for massless particles.
The \finaldraft{total} derivative of $\cal I$ is therefore
\begin{eqnarray}
\label{fullideriv}
\frac{d{\cal I}}{d\lambda} &=& 
    \frac{\partial{\cal I}}{\partial x^\alpha}\frac{d x^\alpha}{d \lambda}
    + \frac{\partial{\cal I}}{\partial p^\alpha}
   \frac{d p^\alpha}{d \lambda} \nonumber\ , \\
   &=&p^\alpha\frac{\partial{\cal I}}{\partial x^\alpha}
   -\Gamma^\alpha_{\beta\gamma}p^\beta p^\gamma
   \frac{\partial{\cal I}}{\partial p^\alpha} \ .
\end{eqnarray}
This, combined with equation (\ref{radtrans}), yields
\begin{eqnarray}
\label{noray}
  &  &
   -p^\alpha u_\alpha\vert_\lambda
   \left[-\chi_0(x^\beta,\nu){\cal I}+\eta_0(x^\beta, \nu)\right] \nonumber\\
  &  &\hspace{3cm}=p^\alpha\frac{\partial{\cal I}}{\partial x^\alpha}
    -\Gamma^\alpha_{\beta\gamma}p^\beta p^\gamma
   \frac{\partial{\cal I}}{\partial p^\alpha}\ , 
\end{eqnarray}
  which is the same as that derived by Lindquist (1966)
  from the Boltzmann Equation. 

The metric and the initial conditions define the rays 
  (the photon trajectories in 3D space), 
  and the solution can be obtained by direct integration along the ray. 
For simplicity, we assume the refractive index $n=1$ throughout the medium.  
The solution to equation (\ref{radtrans}) is then  
\begin{eqnarray}
\label{anaray}
{\cal I}(\lambda) &=& {\cal I}(\lambda_0) 
   \exp\left(\int_{\lambda_0}^{\lambda} 
   \chi_0(\lambda',\nu_0) u_\alpha p^\alpha d\lambda'\right)\\
    & &\hspace{-1cm} 
  -\int_{\lambda_0}^{\lambda}
   \exp\biggl(\int_{\lambda'}^{\lambda}\chi_0(\lambda'', \nu_0)
   u_\alpha p^\alpha d\lambda''\biggr)
   \eta_0(\lambda', \nu_0) u_\alpha p^\alpha d\lambda'\ . \nonumber 
\end{eqnarray} 
In the non-relativistic limit, $u_\alpha p^\alpha = 1$, 
  and the equation recovers the conventional form  
  (see Rybicki \& Lightman 1979).  

\section{Particle Trajectories}   

\subsection{Free particles}

To determines the photon trajectories 
  we need to specify the metric of the space-time. 
We consider the Boyer-Lindquist coordinates: 
\begin{eqnarray}  
 d\tau^2 &=& \biggl( 1- {{2Mr} \over \Sigma}\biggr)dt^2 
         + {{4aMr \sin^2\theta} \over \Sigma}dtd\phi 
         - {\Sigma \over \Delta}dr^2 \nonumber \\ 
     & & \hspace{-0.25cm} - \Sigma d\theta^2  
    - \biggl( r^2+a^2 + {{2a^2Mr \sin^2\theta} \over \Sigma} \biggr) 
                  \sin^2\theta d\phi^2 \ , 
\end{eqnarray}  
  where $M$ is the black hole mass, \finaldraft{with the three vector $(r,\theta,\phi$) corresponding to spherical polar coordinates, and defining} $\Sigma = r^2+a^2\cos^2\theta$ 
  and $\Delta = r^2 -2Mr +a^2$. 
The dimensionless parameter $a/M$ specifies the spin of the black hole, 
  with $a/M = 0$ corresponding to a Schwarzschild (non-rotating) black hole 
  and $a/M = 1$ to a maximally rotating Kerr black hole.

The motion of a free particle is described by the Lagrangian:  
\begin{eqnarray}
 {\cal L} & = & \frac{1}{2} \biggl[ 
   -\left(1-\frac{2Mr}{\Sigma}\right)\dot{t}^2 
   - \frac{4aMr\sin^2\theta}{\Sigma}\dot{t} \dot{\phi}  
   + \frac{\Sigma}{\Delta}\dot{r}^2 \nonumber \\ 
   && + \Sigma \dot{\theta}^2
   + \left( r^2+a^2 + \frac{2a^2M r\sin^2\theta}{\Sigma} \right)
      \sin^2\theta \dot{\phi}^2 
   \biggr]  \   
\label{lagkerreqn}
\end{eqnarray} 
  (here $\dot x^{\alpha} = d x^{\alpha}/ d\lambda$).   
The Lagrangian does not depend explicitly on the $t$ and $\phi$ coordinates.
The momenta in the four coordinates are therefore 
\begin{eqnarray}
  p_{\rm t} = \partial{\cal L}/\partial \dot t &=& -E\ ,\\
\label{pr}
  p_{\rm r} = \partial{\cal L}/\partial \dot r &=& \frac{\Sigma}{\Delta} \dot r\ , \\
\label{ptheta}
  p_{\rm \theta} = \partial{\cal L}/\partial \dot \theta &=& \Sigma \dot \theta\ , \\
  p_{\rm \phi} = \partial{\cal L}/\partial \dot \phi &=& L\ .
\end{eqnarray}
  with $E$ being the energy of the particle at infinity 
  and $L$ the angular momentum in the $\phi$ direction. 
The corresponding equations of motion  are 
\begin{eqnarray}
\label{tdot}
  \dot t & = & E + \frac{2 r(r^2+a^2)E-2a L}{\Sigma\Delta}\ ,\\  
  \dot r^2 & = & {\Delta \over \Sigma} 
    \big(H + E \dot t - L \dot \phi- \Sigma \dot \theta^2 \big)\ , \\ 
  \dot \theta^2 & = & {1 \over \Sigma^2} \big( 
      Q + (E^2 + H)a^2  \cos^2\theta - L^2 \cot^2 \theta \big)\ ,  \\ 
\label{phidot}  
  \dot \phi & = & {{2a rE  + (\Sigma - 2 r)L/\sin^2\theta} \over 
    {\Sigma\Delta} } \ ,   
\end{eqnarray}    
  where $Q$ is Carter's constant (Carter 1968), 
  and $H$ is the Hamiltonian, 
  which equals 0 for photons and massless particles and equals $-1$ 
  for particles with a non-zero mass.  
(See Reynolds et al.\ (1999) for more details.) 
For simplicity, we have set the black-hole \finaldraft{rest} mass equal to unity 
  ($M =1$) in the equations above. 
This is equivalent to normalizing the length 
  to the gravitational radius of the black hole 
  (i.e., set $R_{\rm g} \equiv GM/c^2 = 1$),  
  and we will adopt this normalization hereafter.   

There are square terms of $\dot r$ and $\dot \theta$ 
  in two equations of motion.  
They could cause problems 
  when determining the turning points at which $\dot r$ and $\dot \theta$ change sign in the numerical calculations.   
To overcome this, 
  we consider the second derivatives of $r$ and $\theta$ instead.  
From the Euler-Lagrange equation, we obtain     
\begin{eqnarray}
   \ddot  r & = & \frac{\Delta}{\Sigma}
   \bigg\{\frac{\Sigma-2r^2}{\Sigma^2}\dot t^2+\frac{(r-1)
   \Sigma-r\Delta}{\Delta^2}\dot r^2+ r\dot \theta^2 \nonumber \\
   & & \hspace{1cm} +\sin^2\theta
   \left(r+\frac{\Sigma-2r^2}{\Sigma^2} a^2\sin^2\theta\right)\dot \phi^2
     \nonumber \\
     & & \hspace{1cm} -2a\sin^2\theta\frac{\Sigma-2r^2}{\Sigma^2}\dot t \dot \phi   
       +\frac{2a^2\sin\theta\cos\theta}{\Delta}\dot r \dot \theta\bigg\}\ , \\
   \ddot \theta & = & {1 \over \Sigma} \bigg\{ \sin\theta\cos\theta
   \bigg[ \frac{2a^2r}{\Sigma}\dot t^2 
   -\frac{4ar(r^2+a^2)}{\Sigma}\dot t \dot \phi-\frac{a^2}{\Delta}\dot r^2\nonumber \\
   & & \hspace*{1cm}+a^2\dot \theta^2
   +\frac{\Delta +2r(r^2+a^2)^2}{\Sigma^2}\dot \phi^2 \bigg] 
   -2r\dot r\dot \theta \bigg\}\  .
\end{eqnarray}
In terms of the momenta and the Hamiltonian, the equations above can be expressed as
\begin{eqnarray}
\label{rdot}
   \dot p_{\rm r} &=& 
   \frac{1}{\Sigma \Delta} \left[(r-1)\left((r^2+a^2)H-\kappa\right)
   +r\Delta H\right.\nonumber\\
  & &\hspace{0.5cm}\left.+2r(r^2+a^2)E^2-2aEL\right]-\frac{2{p_r}^2(r-1)}{\Sigma}\ ,\\
\label{thetadot}
  \dot p_{\rm \theta} &=& \frac{\sin\theta\cos\theta}{\Sigma}
   \left[\frac{L^2}{\sin^4\theta}-a^2(E^2+H)\right]\ ,
\end{eqnarray}
  where $\kappa = Q+L^2+a^2(E^2+H)$. 
Equations (\ref{pr}), (\ref{ptheta}), (\ref{tdot}), (\ref{phidot}), 
  (\ref{rdot}) and (\ref{thetadot}) 
  are the equations of motion. 

\subsection{Motion in the presence of external forces} 

The equations of motion obtained in the previous section 
  are applicable to free particles only. 
In a general situation external (non-gravitational) forces may be present 
  and we need to specify the external force explicitly 
  in deriving the equations of motion.  
However, in the setting of accretion disks around black holes     
  we can often treat the effect of the external force implicitly 
  which we will discuss in more detail in the following subsections. 

\subsubsection{Rotational and Pressure Supported Model} 

Here we consider a simple model such that
\begin{equation}
   \dot{t} > \dot{\phi} \gg \dot{r} \gg \dot{\theta}\ .
\end{equation} 
As $\dot{r}$ and $\dot{\theta}$ are small 
  in comparison with other quantities, 
  they can be neglected as a first approximation.

The equation of motion reads  
\begin{equation}
  \frac{d^2 x^\nu}{d\lambda^2}
  +\Gamma^{\nu}_{\alpha \beta}u^{\alpha}u^{\beta}=a^{\nu}\ ,
\end{equation}
  where $a^{\nu}$ is the four-acceleration due to an external force per unit mass.   
For axisymmetry 
  (which is a sensible assumption for accretion onto rotating black holes),   
  $d/d\phi=0$ and $a^{\phi}=0$.  
The identity $u^\alpha a_\alpha = 0$ 
  together with $\dot r = 0$ and $\dot \theta=0$   
  imply that $a^{t}=0$.   
We may also set $a^{r}=0$ for simplicity.  
Because we have an extra equation 
  from the identity $u^{\alpha}u_{\alpha}=-1$, 
  $a^{\theta}$ can be determined self-consistently 
  under the approximation $\dot{\theta}=0$.  
This scenario thus corresponds to flows 
  supported by rotation in the $\hat{r}$ direction 
  and by pressure in the $\hat{\theta}$ direction.  

Inserting the affine connection coefficients for the Kerr metric 
  into the equation of motion 
  yields quantities identical to zero 
  on the left hand side of the equations
  for the $\hat{t}$ and $\hat{\phi}$ directions. 
This leaves only the non-trivial momentum equation 
  in the radial direction: 
\begin{eqnarray}
   0&=&-\left(\frac{\Sigma-2r^2}{\Sigma^2}\right)\dot{t}^2
    +2\left(\frac{\Sigma-2r^2}{\Sigma^2}\right)
    a\sin^2\theta\dot{t}\dot{\phi}\nonumber\\
  &&\hspace{0.5cm}- 
   \left(r+a^2\sin^2\theta\left(\frac{\Sigma-2r^2}{\Sigma^2}\right)\right)
   \sin^2\theta\dot{\phi}^2\ ,
\end{eqnarray}  
   which further simplifies to
\begin{equation}
  \frac{r\Sigma^2\sin^2\theta}{2r^2-\Sigma}\dot{\phi}^2
  =\left(\dot{t}-a\sin^2\theta\dot{\phi}\right)^2 \ .
\end{equation} 
Here we choose the positive solution
\begin{equation}
\label{flowtphi}
   \dot{t}=\left(\frac{\sqrt{r}\Sigma\sin\theta}
   {\sqrt{2r^2-\Sigma}}+a\sin^2\theta\right)\dot{\phi} \ ,
\end{equation}  
  which corresponds to the same rotation as the black hole. 
This solution thus allows the flow to match the rotation of a prograde accretion disk. 

From the metric we have
\begin{eqnarray} 
\label{mertic1}
   1&=&\left(1-\frac{2r}{\Sigma}\right)\dot{t}^2
   +\frac{4ar\sin^2\theta}{\Sigma}\dot{t}\dot{\phi}\nonumber\\
   &&\hspace{0.5cm}-\left(r^2+a^2+\frac{2a^2r\sin^2\theta}{\Sigma}\right)
   \sin^2\theta\dot{\phi}^2 \ .
\end{eqnarray}  
Combining equations (\ref{flowtphi}) and (\ref{mertic1}) 
  yields 
\begin{equation}
    \Sigma\sin^2\theta\left(\frac{r(\Sigma-2r)}{2r^2-\Sigma}
    +\frac{2\sqrt{r}a\sin\theta}
    {\sqrt{2r^2-\Sigma}}-1\right)\dot{\phi}^2=1 \ .
\end{equation} 
It follows that the components of the four-velocity are 
\begin{eqnarray}
\label{mediummotion}
   \dot{t}&=&\frac{1}{\zeta}
   \left(\Sigma\sqrt{r}+a\sin\theta \sqrt{2r^2-\Sigma}\right)\ , \nonumber\\
   \dot{r}&=&0\ ,   \nonumber\\
   \dot{\theta}&=&0\ ,\nonumber\\
   \dot{\phi}&=&\frac{\sqrt{2r^2-\Sigma}}{\zeta\sin\theta} \  ,
\end{eqnarray}
  where 
\begin{equation}
\label{zeta}
   \zeta=\sqrt{\Sigma\left(\Sigma(r+1)
   -4r^2+2a\sin\theta\sqrt{r(2r^2-\Sigma)}\right)} \ . 
\end{equation}

The marginally stable orbit for particles is defined by the surface where
\begin{equation} 
\label{margin_stable}
  \frac{\partial E}{\partial r} = 0 \ .
\end{equation}
From equations (\ref{tdot}), (\ref{phidot}) and (\ref{mediummotion}), 
  we have 
\begin{equation}
  E = \frac{1}{\zeta}\left((\Sigma-2r)
  \sqrt{r}+a\sin\theta\sqrt{2r^2-\Sigma}\right) \ .
\end{equation}
After differentiation, 
  we remove the non-zero factors in the expression and obtain the condition
\begin{equation}
\label{marg_stable}
  \Delta\Sigma^2-4r(2r^2-\Sigma)
  \left(\sqrt{2r^2-\Sigma}-a\sin\theta\sqrt{r}\right)^2=0\ .
\end{equation}
Setting $a=0$ gives $r=6$, 
  which is often regarded as the limit for the inner boundary 
  of an accretion disk around a Schwarzschild black hole.  
This value is the same as that derived 
  by Bardeen, Press \& Teukolsky (1972) 
  using $\partial^2 p_{\rm r}^2 / \partial r^2 = 0$.  

Before we proceed further, 
  we must note that the expressions for the velocity components 
  in equations (\ref{mediummotion}) 
  hold only for regions ``sufficiently'' far 
  from the black-hole event horizon. 
The approximation that we adopt in the model breaks down 
  when the square root in the denominator approaches zero. 
This occurs at the light circularisation radius $r_{\rm cir}$, 
  which is given by $\zeta = 0$, or equivalently
\begin{equation}
\label{light_circ}
  \Sigma(r+1)-4r^2+2a\sin\theta\sqrt{r(2r^2-\Sigma)}\ 
  \bigg|_{r = r_{\rm cir}} = 0 \ .
\end{equation}
Moreover, the assumption of $\dot{\theta}=0$ is also invalid  
  for radii smaller than the radius of the marginally stable orbit 
  --- the flow is neither rotational nor pressure supported 
  and it follows a geodesic into the event horizon.

\subsubsection{Isobaric Surfaces}

In a stationary accretion flow, 
  the acceleration must be balanced by some forces, 
  e.g.\ the gradient of gas or radiation pressure.  
As the local acceleration $a^\alpha$ can be calculated 
  from the rotation law $\omega(r,\theta)$ 
  we can derive a set of isobaric surfaces 
  when a rotation law is given.   
For a barotropic equation of state of the accreting matter
  the isobaric surfaces coincide 
  with the isopicnic (constant-density) surfaces.
  
The accelerations in the ${\hat r}$ and ${\hat \theta}$ directions are 
\begin{eqnarray}
  -\frac{\Sigma}{\Delta}a^r&=&\frac{\Sigma-2r^2}{\Sigma^2}
  \left(\dot{t}-a\sin\theta\dot{\phi}\right)^2+r 
  \sin^2\theta \dot{\phi}^2\ ,  \\
  -\Sigma a^\theta&=&\sin\theta\cos\theta
  \left[\frac{2r}{\Sigma^2}\left(a\dot{t}-(r^2+a^2)\dot{\phi}\right)^2
  +\Delta\dot{\phi}^2\right]\ .
\end{eqnarray}
The surface of constant acceleration is given by
\begin{equation}
\label{tsurface}
   a_\alpha \frac{dx_{\rm surf}^\alpha}{d\lambda} = 0  
\end{equation} 
  (here and hereafter $dx_{\rm surf}^{\alpha}/{d\lambda} 
   \equiv dx^{\alpha}/{d\lambda}|_{x_{\rm surf}}$).  
The stationary condition implies $d t/{d\lambda}= 0$,  
  and axisymmetry implies $d \phi/{d\lambda}=0$. 
Without losing generality, 
  we can choose $t = \phi = 0$ on the surface. 
Thus, equation (\ref{tsurface}) becomes
\begin{eqnarray} 
\label{tsurface2}
   0 &=&  \frac{\Sigma a^r}{\Delta} \frac{d r_{\rm surf}}{d \lambda}
   + \Sigma a^\theta \frac{d \theta_{\rm surf}}{d \lambda}\ ,\nonumber\\
    &=& \beta_1 \frac{d r_{\rm surf}}{d \lambda}
  + \beta_2 \frac{d \theta_{\rm surf}}{d \lambda}\ ,
\end{eqnarray}
  where 
\begin{eqnarray}
   \beta_1&=&\frac{\Sigma-2r^2}{\Sigma^2}
   \left(\frac{1}{\omega}-a\sin\theta\right)^2+r\sin^2\theta \ , \nonumber\\
   \beta_2&=&\sin\theta\cos\theta
   \left[\Delta+\frac{2r}{\Sigma^2}
   \left(\frac{a}{\omega}-(r^2+a^2)\right)^2\right]\ , 
\end{eqnarray} 
   and $d r_{\rm surf}/d \lambda$ and $d \theta_{\rm surf}/d \lambda$  
   determine the intersection of the isobaric surfaces  
   and the $(r,\theta)$ plane.  
By rescaling equation (\ref{tsurface2}) with a factor of $\sqrt{\Delta/\Sigma}$ 
  and making use of the invariance  
\begin{equation}  
  - \left(\frac{d\tau}{d\lambda}\right)^2 
   = \frac{\Sigma}{\Delta}{\dot  r}^2 + \Sigma {\dot  \theta}^2 \ ,  
\end{equation} 
  we obtain  
\begin{eqnarray}
  \frac{d r_{\rm surf}}{d \lambda'}
    &=& \frac{\beta_1}{\sqrt{\beta_2^2+\Delta\beta_1^2}}\ ,   \nonumber  \\
  \frac{d \theta_{\rm surf}}{d \lambda'}
    &=& \frac{-\beta_2}{\sqrt{\beta_2^2+\Delta\beta_1^2}}\ .   
\label{isobaric} 
\end{eqnarray}  
 
These two differential equations can be solved numerically 
  and yield the isobaric surface as a parametric function of $\lambda'$.

\section{Model Accretion Disks and Tori} 

We now demonstrate using the equations of motion 
  above to construct the emitter models. 
The first is a geometrically thin accretion disk, 
  in which the emitting particles are in Keplerian motion.  
The second is a torus, 
  a 3-dimensional object with non-negligible thickness.   

\subsection{Accretion Disk}

When space-time curvature is important, 
  the Keplerian angular velocity  
  of a test particle around a gravitating object  
  is no longer $\omega_k= r^{-3/2}$,  
  the expression in flat space-time.  
Instead, the Keplerian angular velocity 
  in a plane containing the gravitating object  
  can be obtained by setting $\theta=\pi/2$ 
  in equations (\ref{mediummotion}) and (\ref{zeta}). 
Hence, the components of the four-velocity of the particles 
  in the disk are 
\begin{eqnarray}
\label{diskvel}
  \dot{t}&=&\frac{r^2+a\sqrt{r}}{r\sqrt{r^2-3r+2a\sqrt{r}}}\ , \nonumber \\
  \dot{r}&=&0 \ ,\nonumber \\
  \dot{\theta}&=&0 \ ,\nonumber \\
  \dot{\phi}&=&\frac{1}{\sqrt{r}\sqrt{r^2-3r+2a\sqrt{r}}}\ ,
\end{eqnarray}
and the rotational velocity of a Keplerian accretion disk  
  around a black hole is  
\begin{equation}
  \omega_k=\frac{1}{r^{3/2}+a},
\end{equation}  
  (Bardeen, Press \& Teukolsky 1972).   

The relative energy shift of the emission 
  between the disk particle and an observer at a large distance is 
  determined by equation (\ref{freqshift}), 
  with $u^\alpha$ as given in equation (\ref{diskvel}).  
Keplerian disk images can be found in many existing works 
  (e.g.\  Bromley, Miller \& Pariev 1998), and    
  we do not show disk images here. 
The general characteristics are that a disk image is asymmetric,  
  with the separatrix for the energy shift of the emission 
  no longer bisecting the disk image into two equal sectors, 
  one for red shift and another for blue shift.
The whole disk appears to be reddened, especially at the inner rim.

\subsection{Accretion Torus} 

To determine the geometry and structure of an accretion torus  
  self-consistently is beyond the scope of this paper. 
Here, we consider a simple parametric model, 
  with an angular velocity profile given by 
\begin{equation}
  \omega=\frac{1}{(r\sin\theta)^{3/2}+a}
  \left(\frac{r_{\rm k}}{r\sin\theta}\right)^n \ . 
\label{rk-equation} 
\end{equation}
The quantity $r_{\rm k}$ is the radius (on the equatorial plane)  
  at which the material moves with a Keplerian velocity.  
The parameter $n$ adjusts the force term, such as a pressure gradient, 
  to keep the disk particles in their orbits, 
  and it determines the thickness of the torus.  
In this study we just take $n=0.21$ without losing generality.   
If the torus is supported by radiation pressure, 
  its inner edge is determined by the intersection of the isobaric surface  
  with either one of two surfaces. 
These two surfaces provide the constraints, 
  inside which the pressure-supported solution does not hold.   
The first is a surface defined by the orbits of marginal stability.    
For $\omega(r,\theta)$, it is given by 
\begin{eqnarray}
  0 &=&  2a\sin^4\theta\left[\frac{r^2}{\Sigma}
   -\left(r^2+a^2+\frac{a^2r\sin^2\theta}{\Sigma} \right)
  \frac{\Sigma-2r^2}{\Sigma^2} \right]\omega^3  \nonumber\\
  && +\sin^2\theta
   \bigg[\left(\frac{6r(r^2+a^2)}{\Sigma}+3\Delta-\Sigma\right)
   \frac{\Sigma-2r^2}{\Sigma^2}  \nonumber \\ 
  && \hspace*{0.5cm} +r\left(1-\frac{2r}{\Sigma}\right)\bigg]\omega^2 
   -\frac{6ar\sin^2\theta}{\Sigma}
     \left(\frac{\Sigma-2r^2}{\Sigma^2}\right)\omega \nonumber\\
 &&  +\Delta\sin^2\theta ~\omega\frac{\partial\omega}{\partial r}  
     -\left(1-\frac{2r}{\Sigma}\right)\frac{\Sigma-2r^2}{\Sigma^2} \ . 
\end{eqnarray}  
The second is the limiting surface 
  where the linear velocity approaches the speed of light. 
It is given by  
\begin{eqnarray}
   0 &=& \Sigma-2r+4ar\omega\sin^2\theta  \nonumber\\
   & &\hspace{1cm}
  -\left((r^2+a^2)\Sigma+2a^2r\sin^2\theta \right)\omega^2\sin^2\theta \ . 
\end{eqnarray}  
Usually the former is larger than the latter.
The outermost of these two surfaces determines the inner boundary 
  and hence the critical surface of the torus.   

\begin{figure}[ht]  
\vspace*{0.35cm} 
\center{\epsfig{figure=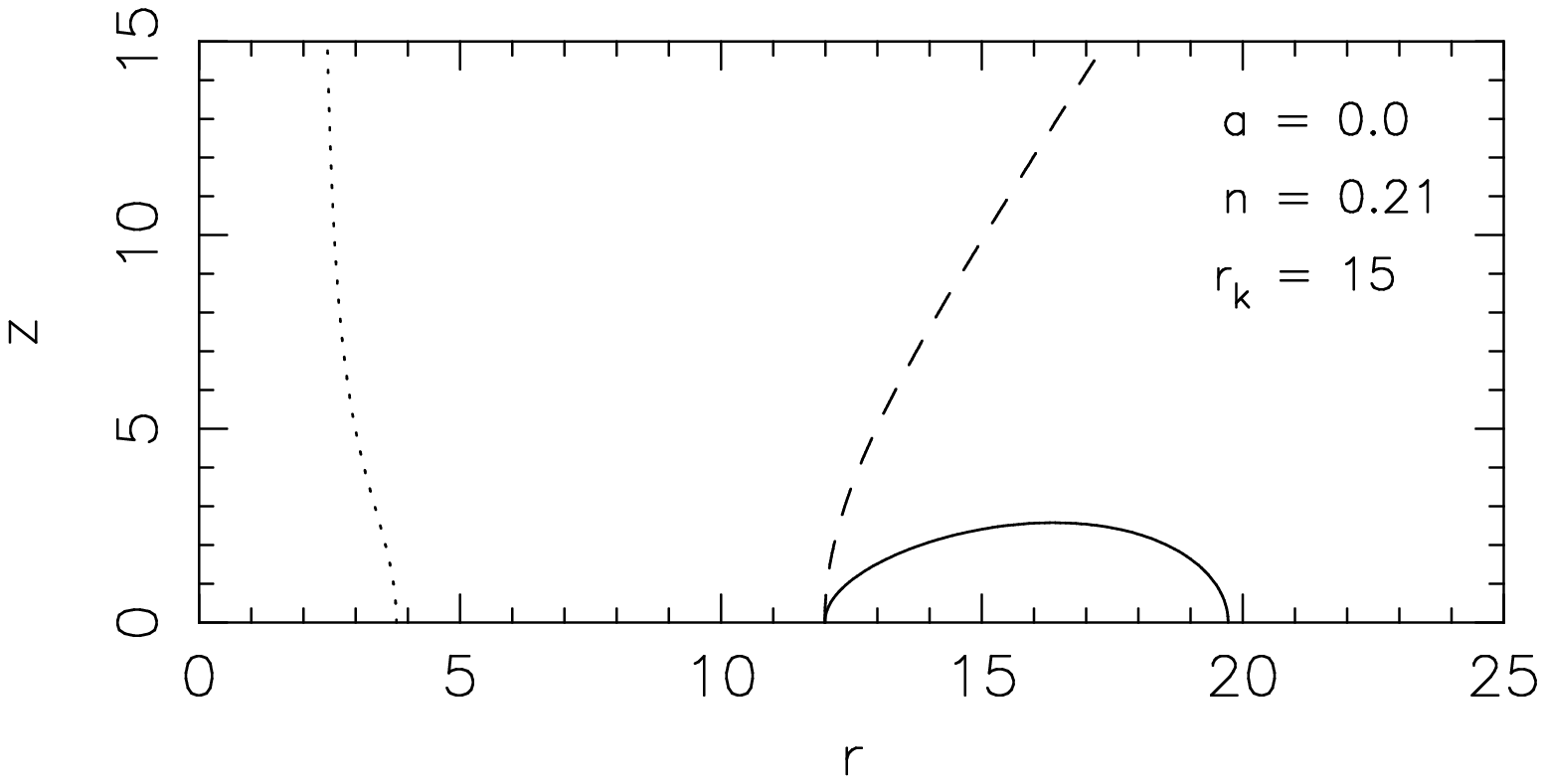,width=8cm}} 
\center{\epsfig{figure=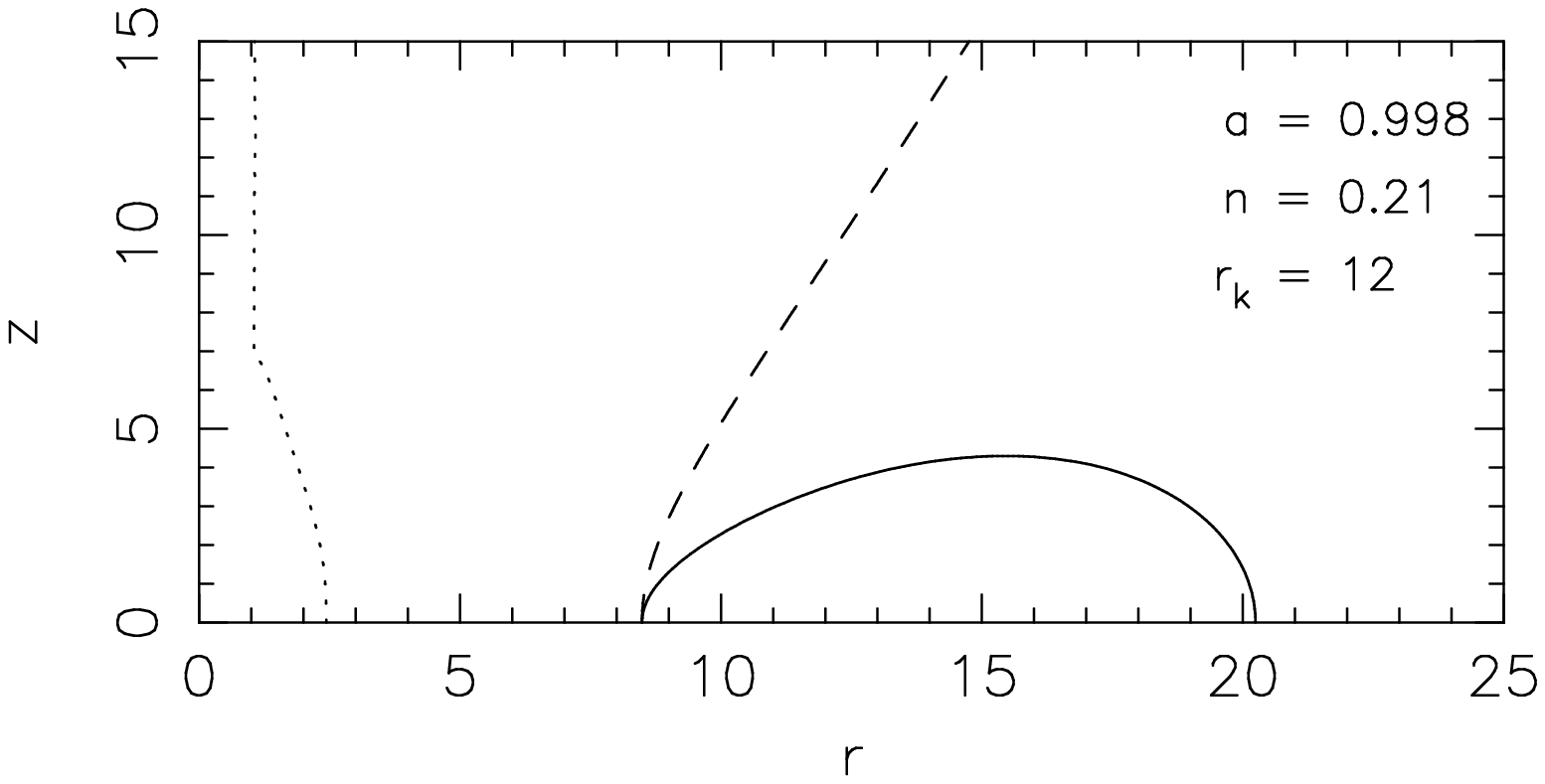,width=8cm}} 
\caption{
  The critical emitting surface of the torus (solid line), 
    the surface defined by the orbits of marginal stability (dashed line)
    and the limiting surface 
    where the linear velocity derived from equation (\ref{rk-equation}) 
    approaches the speed of light (dotted line) 
    for a Schwarzschild black hole (top panel) and a Kerr black hole 
    with $a = 0.998$ (bottom panel). 
  In both cases we consider an profile index $n = 0.21$ 
    (see Equation [\ref{rk-equation}]), 
    which gives the tori an angular momentum profile similar to 
    those obtained in the accretion disk simulation of Hawley \& Balbus (2002).  
  The Kepler radius $r_{\rm k}$ is 15 for the Schwarzschild black hole 
    and is 12 for the Kerr black hole.   
} 
\label{various_surface} 
\end{figure}

Figure \ref{various_surface} shows the critical density surfaces of two tori. 
The first torus is around a Schwarzschild black hole 
  and the second torus is around a maximally rotating black hole. 
The tori are constructed such that their specific angular momentum 
  has a profile similar to those of the simulated accretion disks 
  in Fig. 3. of Hawley \& Balbus (2002).

In our model the boundary surface of the torus is determined
 by a single parameter, $n$,
 which specifies the index of the angular-velocity power law.
Its value is selected
 such that the angular-velocity profile matches
 the profiles obtained by the numerical simulations
 --- here we consider that of Hawley \& Balbus (2002).
Model tori can be constructed using various different methods.
An example is that in a study of dynamical stability of tori
 around a Schwarzschild black hole carried out by Kojima (1986),
 the model parametrizes the angular momentum
 instead of the angular velocity.
We note that the aspect ratios of the torus surfaces
 obtained by Kojima (1986) and those shown in Fig. \ref{various_surface}.
 are similar. 

\section{Extinction}  

\begin{figure}[ht]  
\vspace*{0.5cm}
\center{\epsfig{figure=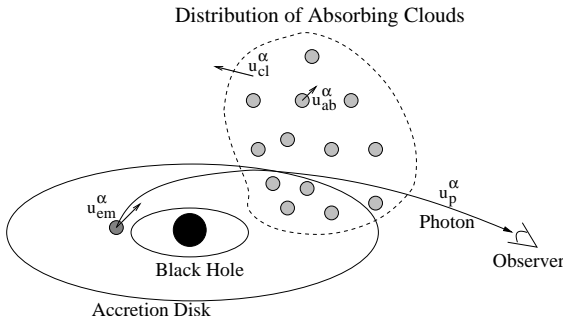,width=7.4cm}}
\vspace*{0.25cm}
\caption{
   A schematic illustration of the physical setting 
     on which the radiative-transfer formulation is constructed.  
   The space-time curvature is determined by a gravitating body, 
     which is a rotating black hole. 
   The emitting object is the accretion disk around the hole. 
   The cloudlets in the vicinity attenuate the radiation 
     from the accretion disk.  
   The properties of the radiation are specified   
     by the 4-velocity of the photon, $u^\alpha_{\rm p}$.   
   The other three important quantities 
     in the model are the 4-velocity of emitting elements 
     in the accretion disk $u^\alpha_{\rm em}$, 
     the bulk velocity of the absorbing cloudlets $u^\alpha_{\rm cl}$ 
     and the ``microscopic'' 4-velocities 
     of the cloudlets $u^\alpha_{\rm ab}$  
     with respect to the bulk motion of the cloudlets. 
}
\label{cloud_model} 
\end{figure}

The generic setting of the system under our investigation  
  is that emitters with various strengths are distributed  
  in space in a curved space-time,  
  and the radiation is attenuated, and may be re-emitted, when propagating. 
The emitters and the line-of-sight material are in relativistic motion 
  with respect to the observer and also with respect to each other.      
An example is that shown in Fig. \ref{cloud_model}., 
  in which the emitters are the surface elements of an accretion disk  
  and the absorbers are some clouds in the vicinity of the disk. 
The photon trajectories and the motion of the emitters and absorbers 
  are affected by the space-time distorted by the central black hole.  

To construct the model we need to determine
\begin{itemize}
\item[(i)] the rays that connect the emitters, absorbers and observer,   
\item[(ii)] the four-velocities of the emitters and absorbers /scatterers, 
\item[(iii)] the spatial distributions 
     of the emitters and the absorbers/scatterers, and  
\item[(iv)] the effective cross section of the absorbers/scatterers.
\end{itemize}
In the previous section, we have shown how to obtain (i) and (ii);  
  in this section we incorporate (iii) and (iv)  
  into the radiative-transfer calculations.    
  
\subsection{An illustrative model}   

We consider a model with the geometry 
  shown in Fig. \ref{cloud_model}.  
The photons are emitted from the elements on the top and bottom surfaces 
  of a geometrically thin disk in a Keplerian rotation 
  around a Kerr black hole.  
The radiation is resonantly scattered (absorbed) by plasma clouds 
  and is attenuated in its propagation.
\finaldraft{We omit emission from the clouds (i.e. $\eta_0 =0$ in equation [\ref{radtrans}]) in this study.}
The size of the clouds is small 
  in comparison with the length scale of the system. 
They are not confined to be in the equatorial plane 
  and are in orbital motion, supported by some implicit forces  
  (which may be radiation, kinematic or magnetic pressure gradients).   
These clouds have a large (thermal) distribution of velocities,
  in addition to their collective bulk velocity.

We assume that the radiation scattered into the energies of the lines 
  is insignificant    
  and ignore the photons that are scattered into the line-of-sight. 
Under this approximation, scattering simply removes the line photons     
  and causes extinction similar to true absorption. 
Thus, for simplicity, hereafter we do not distinguish 
  between scattering and absorption, \footnote{\finaldraft{Here, in spite of the term ``resonant scattering'' (which describes the nature of the process), we have ignored the scattering of continuum photons into the line energy bins and the energy (frequency) redistribution in our calculations.}} 
  and the two terms are interchangeable, 
  unless otherwise stated explicitly.      

The rays originating from the accretion-disk surface 
  that can reach the observer 
  are determined by the 4-momenta of the photons, 
  which are calculated 
  using equations (\ref{tdot}), (\ref{phidot}) and (\ref{rdot}).  
The 4-velocities of the emitting surface elements on the accretion disk  
  are given by equations (\ref{diskvel}). 
These determine the relative energy shifts of the photons   
  between the emitters and the observer.  
What we need next is to determine the relative energy shifts  
  between the emitters and the absorbers. 
Then, we need a model mechanism by which the absorption takes place, 
  and to derive the resonant absorption condition 
  for the absorption coefficient.   

Now we construct a model 
  for the spatial distribution and the velocities of the absorbers. 
Consider a parametric model in which the bulk 4-velocities of the clouds  
  are given by equations (\ref{mediummotion}) and (\ref{zeta}).  
In this model the bulk velocities of the clouds in the equatorial plane
  matches the 4-velocities of the accretion disk.  

The clouds themselves are cold, and the thermal velocity of the gas particles inside are much smaller than their bulk motion and root-mean-square velocity dispersion.
However, the clouds have a large velocity dispersion, given by the local virial temperature, which is comparable to the energy of the emission lines of interest.
Therefore, the clouds can be considered as relativistic particles 
  in the calculation.  
Using the bulk-motion velocities 
  obtained by (\ref{mediummotion}) and (\ref{zeta}) 
  together with the virial theorem, 
  we can derive this temperature 
  and determine the velocity distribution 
  of the absorbing clouds.   

The clouds fill most of space, 
  with a radially dependent number density.  
However, close to the black hole, the assumptions above break down, 
  and the axial force cannot support the clouds out of the equatorial plane.  
When this happens, they will flow along geodesics directly into the hole.  
In the numerical calculation, 
  we determine the asymptotic boundaries 
  at which the left hand sides of equations (\ref{light_circ}) and (\ref{marg_stable}) 
  vanish.  
This is done by evaluating these expressions and testing 
  to see if they are negative at each point along the photon rays.  
Inside that surface, 
  the number density of the clouds will be much less than that outside, 
  which we approximate by setting it to zero in this zone. 

\subsection{The absorption coefficient}  

We assume that the absorption is due to ``cold'' cloudlets 
   with high virial velocities.  
The absorption coefficient of individual cloudlets is  
\begin{eqnarray} 
 \chi_{\rm i} & \propto & 
  \sigma ~ \delta\left(\frac{u^\alpha k_\alpha+\Eline}{\Egamma}\right)    
\end{eqnarray}  
 (with $\sigma$ as the effective absorption cross section of the cloudlet, and $k_\alpha$ the photon four-momentum).  
 The absorption rest frequency is $\Eline$, and $\Egamma$ is the energy of the     photon in the bulk rest frame.    
The total effective absorption coefficient $\chi_0$
  is the sum of the contribution of these cloudlets, i.e.,  
\begin{eqnarray}  
  \chi_0 & = & \sum_{\rm i} ~\chi_{\rm i} \ .   
\end{eqnarray}  
Converting the sum into an integral in momentum space
  yields the absorption per unit length in the rest frame as
\begin{eqnarray}
\label{chieqn}
  \chi_0&=&\frac{-2\pi\lambda\sigma}{\Eline^2}\times  \nonumber\\
&&\!\!\!\int\!\!\!\int\!\! p^2dpd\mu \exp(-E/\Theta) u^\alpha k_\alpha 
  \delta\left(\frac{u^\alpha k_\alpha + \Eline}{\Egamma}\right),
\end{eqnarray}
where we have defined $\mu = \cos\theta$, 
$\lambda$ is a normalization constant, 
$\Theta$ is the temperature in relativistic units (with $k_B=1$), 
and $E$ and $p$ are the energy and momentum of a gas particle in the bulk rest frame.  This form assumes isotropic thermal motion in the rest frame. 

There are three terms that we need to determine  
  before we can evaluate the integral, 
  and they are the normalisation parameter $\lambda$, 
  the temperature $\Theta$ 
  and the photon energy in the rest frame of the absorbing particle 
  $u^\alpha k_\alpha$. 
When these variables are determined, 
  we can then parametrise $\sigma$ after integration is carried out.

\subsubsection{The normalisation parameter $\lambda$}

We now derive $\lambda$ starting from
\begin{equation}
N=4\pi\lambda\int^\infty_0 p^2dp \exp\left(-E/\Theta\right)\ ,
\end{equation}
where $N$ is the number density of absorbing clouds.  
Integrating yields
\begin{equation}
\lambda=\frac{N\frac{m}{\Theta}}{4\pi m^3K_2(\frac{m}{\Theta})}\ ,
\end{equation}
where $K_\nu(x)$ is a modified Bessel function, and $m$ is the average cloud mass.
This is called the J\"{u}ttner distribution 
  and corresponds to Maxwell's distribution 
 except in the case of a relativistically high temperature.

\subsubsection{The temperature $\Theta$}

The total energy in the distribution of clouds is given by
\begin{equation}
E_{\rm tot}=4\pi\lambda\int^\infty_0 p^2dp E \exp\left(-E/\Theta\right)\ .
\end{equation}
Integrating this yields the energy per unit mass as
\begin{equation}
\label{Etot2}
\frac{E_{\rm tot}}{Nm}=\frac{K_3(\frac{m}{\Theta})}
{K_2(\frac{m}{\Theta})}-\frac{\Theta}{m}\ .
\end{equation}

Using conservation of energy and angular momentum, 
we calculate the thermal energy of the virialised relativistic gas 
of absorbing clouds. 

 At infinity the medium has
\begin{eqnarray}
  E_{\rm init}&=&Nm \ ,  \\
  L_{\rm init}&=&L_{\rm fin} \ .
\end{eqnarray}
Close to the black hole it has
\begin{eqnarray}
E_{\rm fin} &=&\frac{Nm}{\zeta}\left[(\Sigma-2r)\sqrt{r}
   +a\sin\theta\sqrt{2r^2-\Sigma}\right]\ ,   \\
L_{\rm fin}&=&\frac{Nm}{\zeta}\left[2ar\sqrt{r}\sin^2\theta
  -(r^2+a^2)\sin\theta\sqrt{2r^2-\Sigma}\right]
\end{eqnarray}

The energy released by the gas falling from infinity 
   and slamming into a wall moving 
   with a velocity given by equation (\ref{mediummotion}) is
\begin{equation}
-E_{\rm tot}=u^\alpha p_\alpha
   =-E_{\rm fin}\dot{t}_{\rm init}-L_{\rm fin}\dot{\phi}_{\rm init}\ .
\end{equation}
After simplification, this becomes
\begin{eqnarray}
\label{Etot1}
-\frac{E_{\rm tot}}{Nm}\!&=&\!\frac{1}{\zeta^2}
\left[(2r^2-\Sigma)(r^2+a^2)
   -2ar\sin\theta\sqrt{r}\sqrt{2r^2-\Sigma}\right. \nonumber\\
&&\hspace{0.5cm}\left.-\left(a\sin\theta\sqrt{2r^2-\Sigma}
   +\Sigma\sqrt{r}\right)\zeta\right]\ .
\end{eqnarray}

Thus the temperature of the media can be derived using 
equations (\ref{Etot2}) and (\ref{Etot1}).  
Unfortunately, this yields an implicit relation 
of $m/\Theta$ that contains transcendental functions.  
The modified Bessel functions can be expanded in the limit 
where $\Theta \ll m$ which corresponds to an ``almost relativistic'' gas.  
Since the potential energy released in accretion is of the order of a 
few percent of the rest mass of the infalling material, this approximation should hold in AGN.

Expanding to second order in $\Theta / m$, cancelling the exponential factors, and then solving the resulting quadratic yields
\begin{equation}
\frac{\Theta}{m} 
  = \frac{2}{5}\left(-1+\sqrt{1+\frac{10}{3}
  \left(\frac{E_{\rm tot}}{Nm}-1\right)}\right)\ .
\end{equation}
Thus we have an explicit description of how the kinematic temperature varies with position.

\subsubsection{The photon energy $u^\alpha k_\alpha$} 
 
In the rest frame, the motion of the thermalised medium is isotropic. 
Thus we can simplify the problem  by aligning an axis 
  along the photon propagation vector and working in a local Lorentz frame so that
\begin{equation}
  k_\alpha = \Egamma(-1, 1, 0, 0)\ ,
\end{equation}
and
\begin{equation}
  p^\alpha = m u^\alpha =(E, p \mu, p_{\rm y}, p_{\rm z})\ .
\end{equation}
Thus, 
\begin{equation}
u^\alpha k_\alpha = \frac{\Egamma}{m}(p\mu - E)\ .
\end{equation}

\subsubsection{Evaluation of the $\delta-$function}

Using the relation that
\begin{equation}
\frac{d}{d\mu}\ \left[\frac{u^\alpha k_\alpha + \Eline}{\Egamma}\right] = -\frac{p}{m} \ ,
\end{equation}  
we obtain  
\begin{eqnarray}
\int d\mu \ u^\alpha k_\alpha 
  \delta\left(\frac{u^\alpha k_\alpha + \Eline}{\Egamma}\right) 
   & & \nonumber \\
   & & \hspace{-2cm} = 
     {\frac{u^\alpha k_\alpha}{\big\vert \frac{d}{d\mu}\left[\frac{u^\alpha k_\alpha + \Eline}{\Egamma}\right]\big\vert}}\bigg\vert_{u^\alpha k_\alpha =- \Eline} \nonumber \\
      & & \hspace{-2cm} = -\Eline \frac{m}{p}\bigg\vert_{u^\alpha k_\alpha =- \Eline}  \ . 
\end{eqnarray}
We change the variable $p dp$ to $EdE$. 
The integral in equation (\ref{chieqn}) can now be simplified to
\begin{equation}
\chi_0=\frac{2\pi\lambda \sigma m}{\Eline}\int^\infty_{\frac{m}{2}\left(\frac{\Egamma}{\Eline}+\frac{\Eline}{\Egamma}\right)} EdE \exp(-E/\Theta)
\end{equation}
Integrating this gives the absorption coefficient in the rest frame of the gas
\begin{eqnarray}
\label{abschi0}
\chi_0&=&\frac{N\sigma}{2 K_2(\frac{m}{\Theta})}\left[\frac{1}{2}\left(\frac{\Egamma}{\Eline}+\frac{\Eline}{\Egamma}\right)+\frac{\Theta}{m}\right]\times\nonumber\\ &&\hspace{1.5cm}\exp\left[-\left(\frac{m}{2\Theta}\right)\left(\frac{\Egamma}{\Eline}+\frac{\Eline}{\Egamma}\right)\right]
\end{eqnarray}
This equation can be recast in terms of $\chi$ by using equations (\ref{chframe}) and (\ref{freqshift}).

\subsubsection{The effective absorption $N\sigma$} 

The absorption coefficient depends upon $N$, the number density of the clouds, and on $\sigma$, the absorption cross-section per cloud.  These are in general a function of position.  Since the pressure gradient and the inflow velocity in the $\hat{r}$ and $\hat{\theta}$ directions are ignored in our approximation of the flow, the density profile cannot be determined in a fully self-consistent manner via the mass continuity equation.  To overcome this, we assume a simple two-parameter profile
\begin{equation}
N\sigma=\sigma_0 r^{-\beta}\ .
\end{equation}
Where $\sigma_0$ can be considered as a proportionality constant fixing the density and opacity scales.  Without losing generality, we adopt a value $\beta = 3/2$.  The results can then easily be generalized to other values of $\beta$.   
For the case of a swarm of absorbing clouds around a black hole, the optical depth is approximately one.  This corresponds to the case where $\sigma_0$ is of order $0.5$ or greater, with the integration proceeding radially to the event horizon.  The effective optical depth depends greatly on the paths the photons take. 

\section{Spectral Calculations}    
 
\subsection{No absorption}   

\subsubsection{Accretion disks} 
\begin{figure}
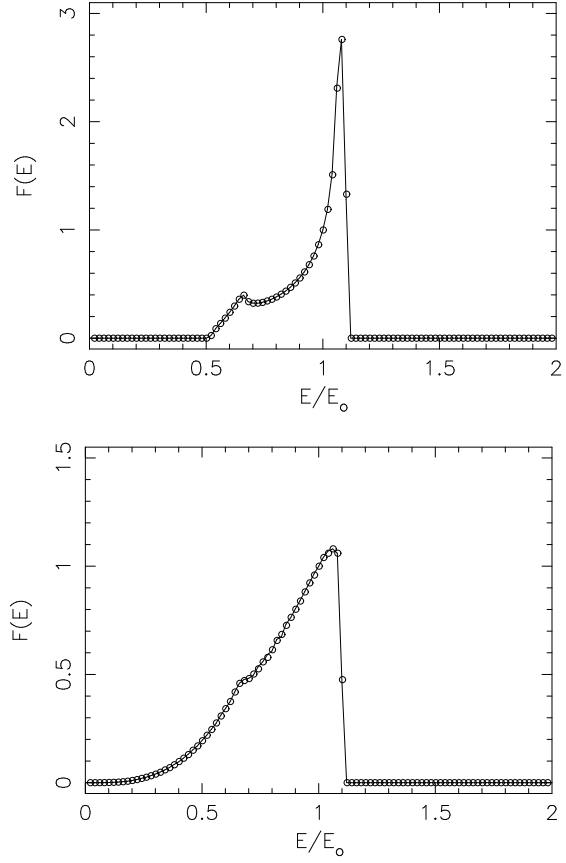

\center{
\epsfig{figure=0814f04.ps,   width=7.25cm} 
\epsfig{figure=0814f05.ps, width=7.25cm}
}
\caption{
  The profiles of emission lines from thin Keplerian accretion disks 
    around a Schwarzschild black hole (top) 
    and a Kerr black hole with $a=0.998$ (bottom). 
  The line profiles are normalized such that the flux $F(E) = 1$ at $E/E_0 =1$. 
  The viewing inclination angle is $45^{\circ}$.  
  The inner radius of the accretion disk is at
    the last stable orbit, 
    and the outer radius is 10$R_{\rm g}$.   
  The line emissivity on the disk surface 
    is a power-law which decreases radially outward, 
    and the powerlaw index is $-3$.  
  Only emission contributed by the direct disk image is considered; 
    emission from higher-order images is not included. 
  The emission is unabsorbed.  
  In each case, the solid lines correspond to the line profiles 
    obtained by semi-analytic calculations described 
    in Fanton et al.\ (1997), 
    and the circles represent the line profile obtained 
    by our numerical ray-tracing calculations.    
}
\label{fantonfig}
\end{figure} 

\begin{figure*}[ht] 
\vspace*{0.5cm}
\center{
\epsfig{figure=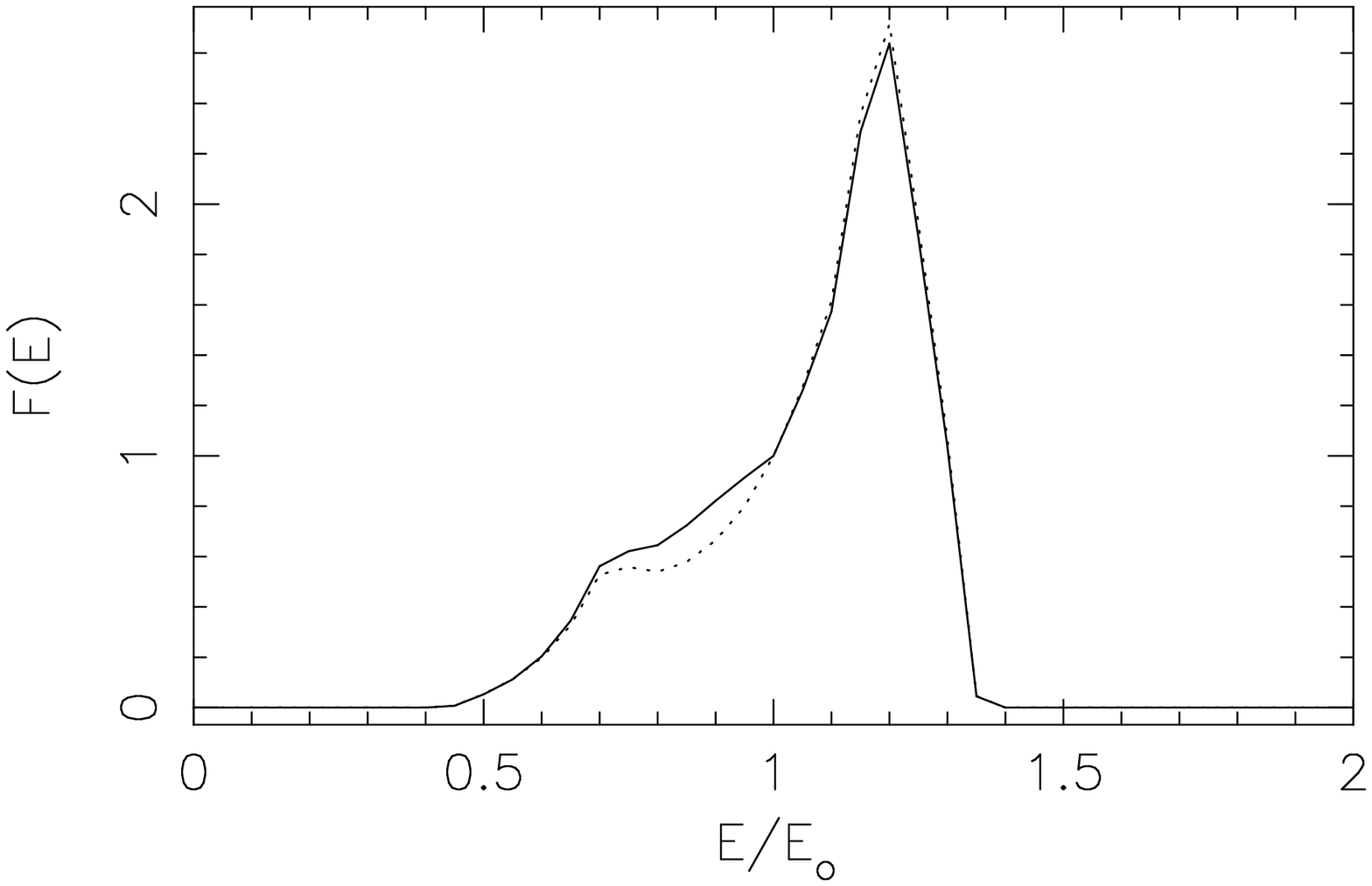, width=7.8cm} 
\epsfig{figure=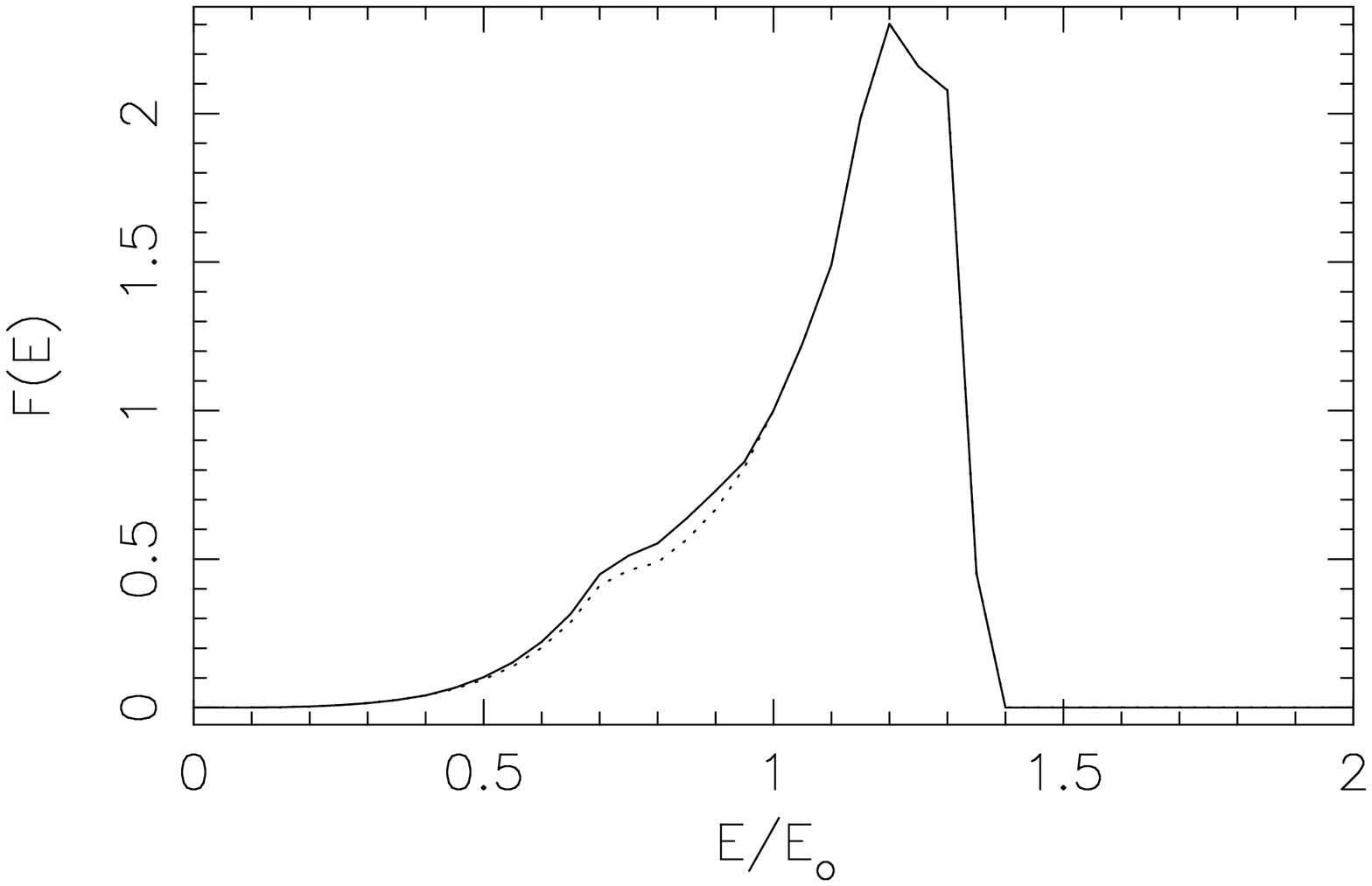, width=7.8cm}    
\epsfig{figure=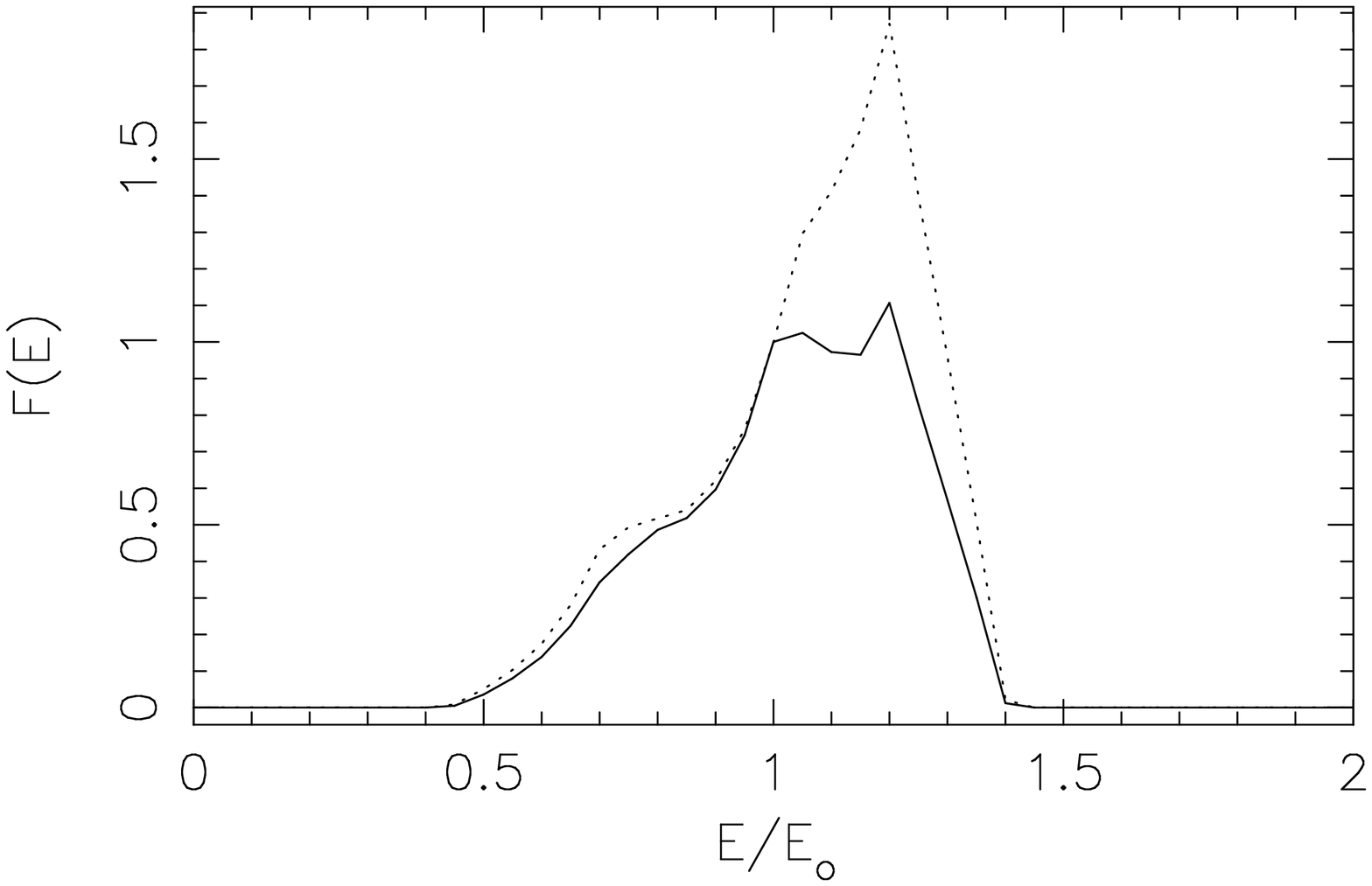, width=7.8cm}
\epsfig{figure=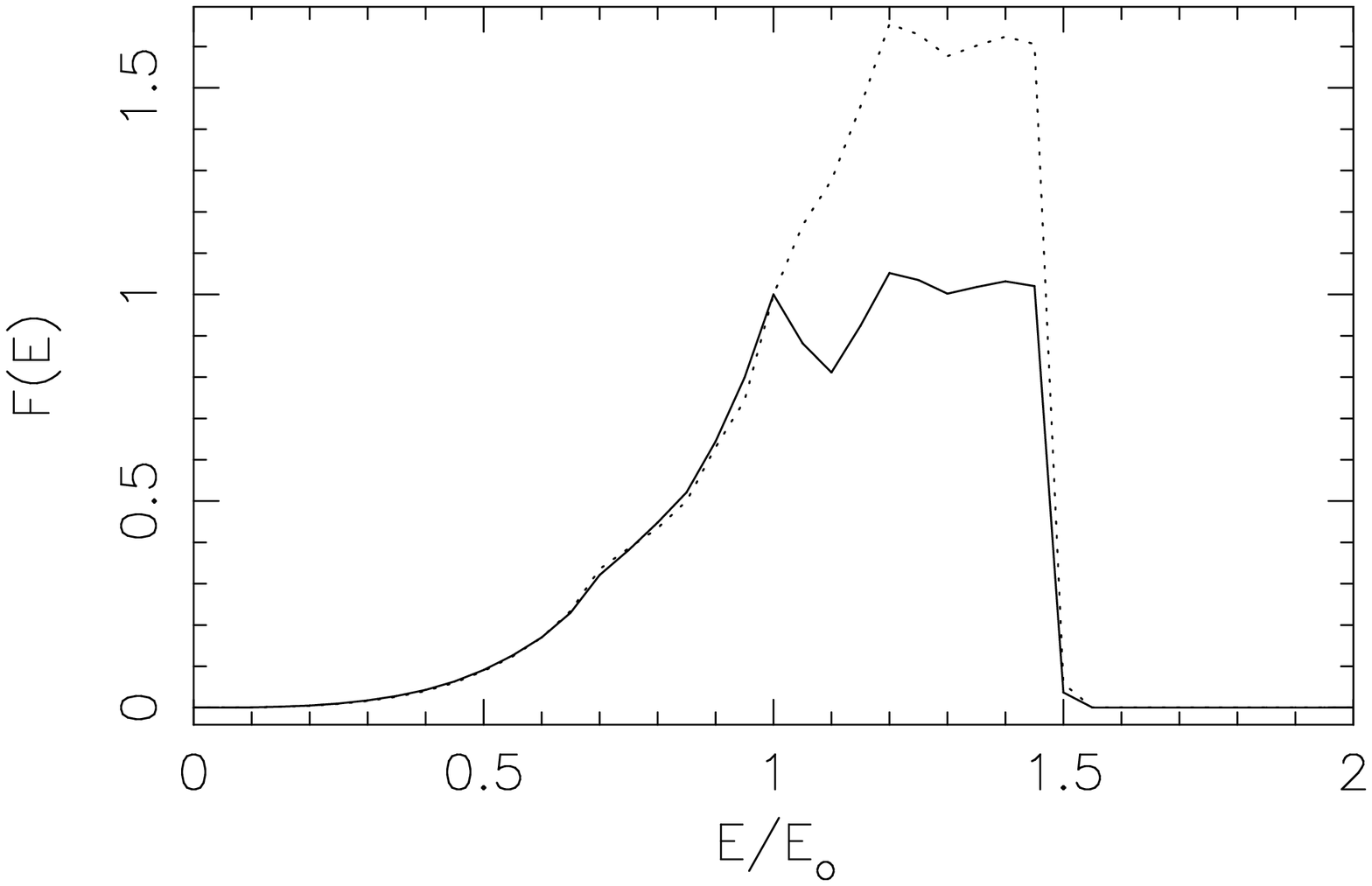, width=7.8cm}   
\epsfig{figure=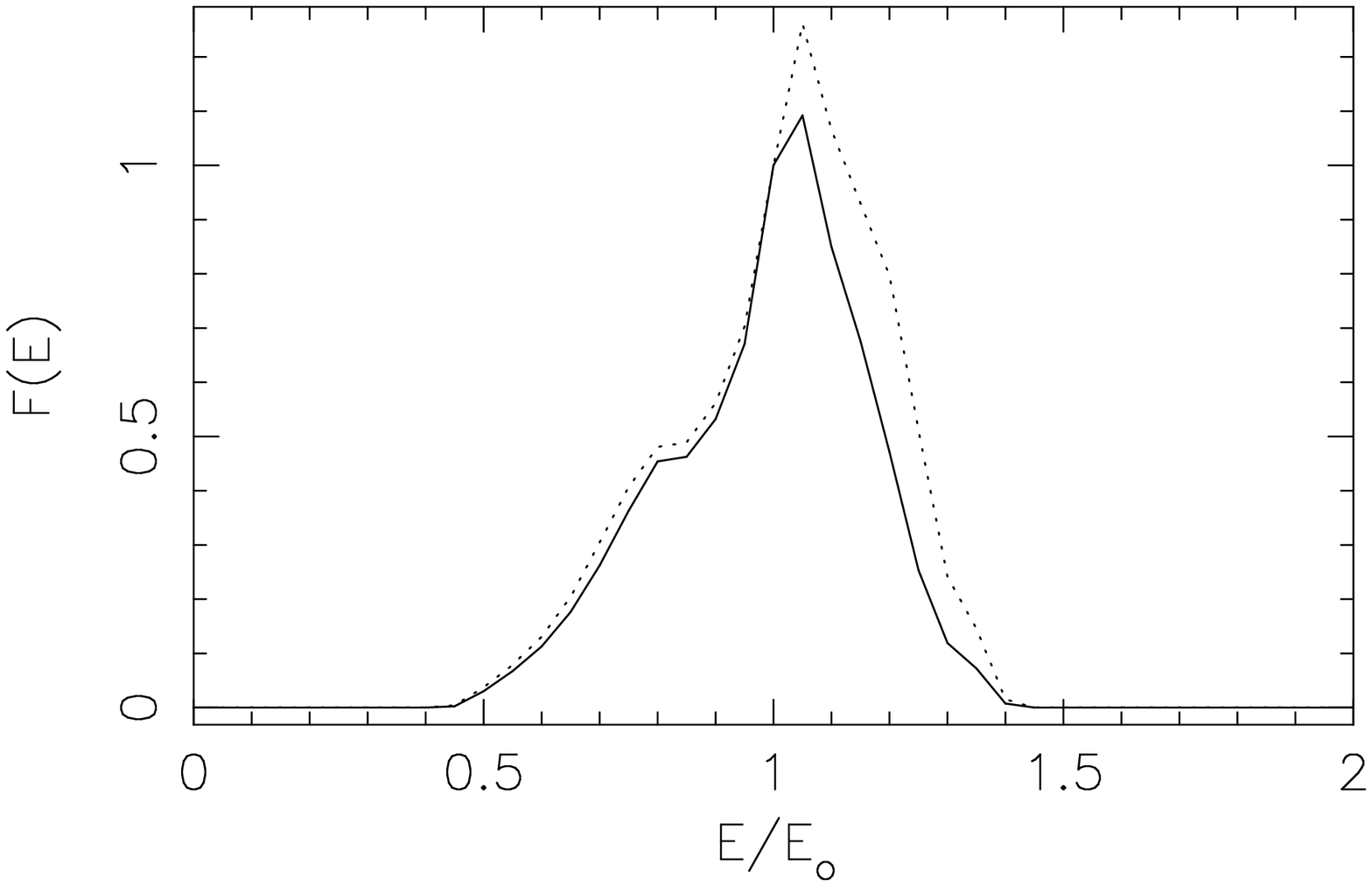, width=7.8cm} 
\epsfig{figure=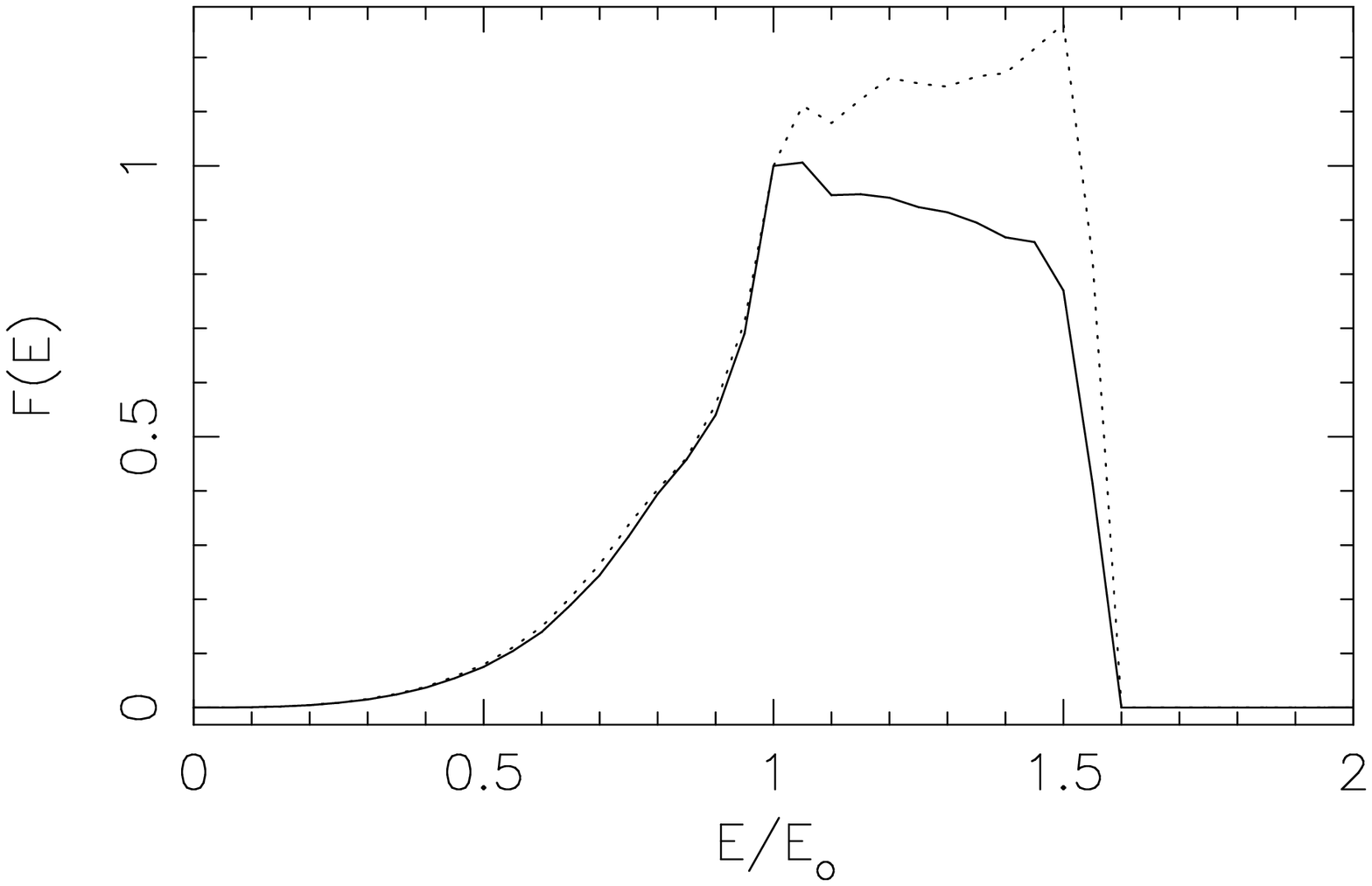, width=7.8cm}
}
\caption{ 
  Comparison of the line profiles 
     for cases considering only the direct disk image (dotted line)
     and cases including high-order images (solid line). 
  The line profiles are normalized such that $F(E) = 1$ at $E/E_0 =1$.
  The viewing inclination angles of the disks are 
     75$^{\circ}$, 85$^{\circ}$ and 89$^{\circ}$ 
     (panels from top to bottom).  
  Line profiles for Schwarzschild black holes are shown in the left column; 
     line profiles for the Kerr black holes with $a = 0.998$ 
     are shown in the right column. 
  The outer radius of the accretion disk is 20$R_{\rm g}$, 
     and the index of the emissivity powerlaw is $-2$.    
}
\label{multi_orders}
\end{figure*}
\vspace*{0cm}

We calculate the observed energy of the flux from a point 
  on the planar disk using equation (\ref{freqshift}).   
We ignore absorption.  
The line emissivity has a power law profile, 
  which decreases radially from disk centre. 
The intensity is proportional to the third power of the relative shift; 
  the flux has an extra factor of $\nu_0/\nu$ 
  due to time dilation, 
  i.e.\ it is proportional to the fourth power 
  of the relative frequency shift. 
(Note that $F$ in \S 2 is the distribution function, 
  not the flux of the emission.) 

Our calculation reproduces the line profiles 
  of direct images of accretion disks 
  as those shown in Fabian et al (1989),  Kojima (1991),
  Fanton et al (1997), Bromley et al (1997), and Reynolds et al (1999). 
Figure \ref{fantonfig} shows two example line spectra 
  calculated using the method described above. 
The spectra contain only emission from the direct image.  
We also show line \finaldraft{spectra} obtained 
  using the method by Fanton et al (1997) for comparison.   
The results in the two ca lculations are in excellent argeement.

We also carry out spectral line calculations 
  which include contribution from higher-order disk images.  
(Here and hereafter   
  we assume that the emissivity powerlaw  
  has an index of $-$2, 
  except where otherwise stated explicitly. \finaldraft{i.e. ${\cal{I}}(\lambda_0)\propto r^{-2}$})    
Our calculations show that  
  the contribution of the higher-order images 
  are significant only at high inclination angles
  (see Fig.\ \ref{multi_orders}). 
The emission from high-order images 
  is mostly at frequencies close to the rest frequency of the line, 
  because the region 
  where highly red and blue-shifted emission originates 
  is obscured.
  
\begin{figure}
\vspace*{0.25cm} 
\center{
\epsfig{figure=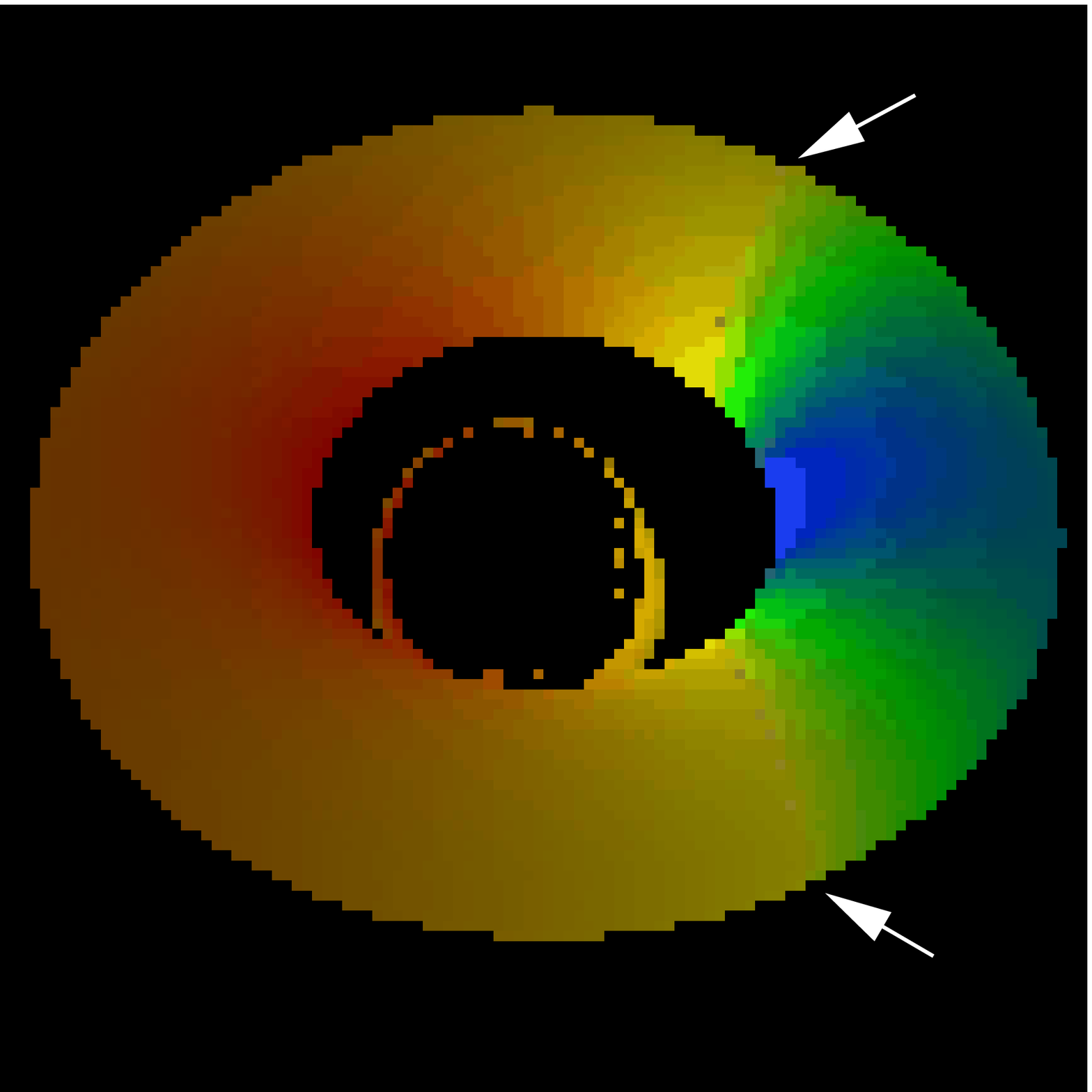, width=7.5cm}
} 
\vspace*{0.1cm}  
\center{
\epsfig{figure=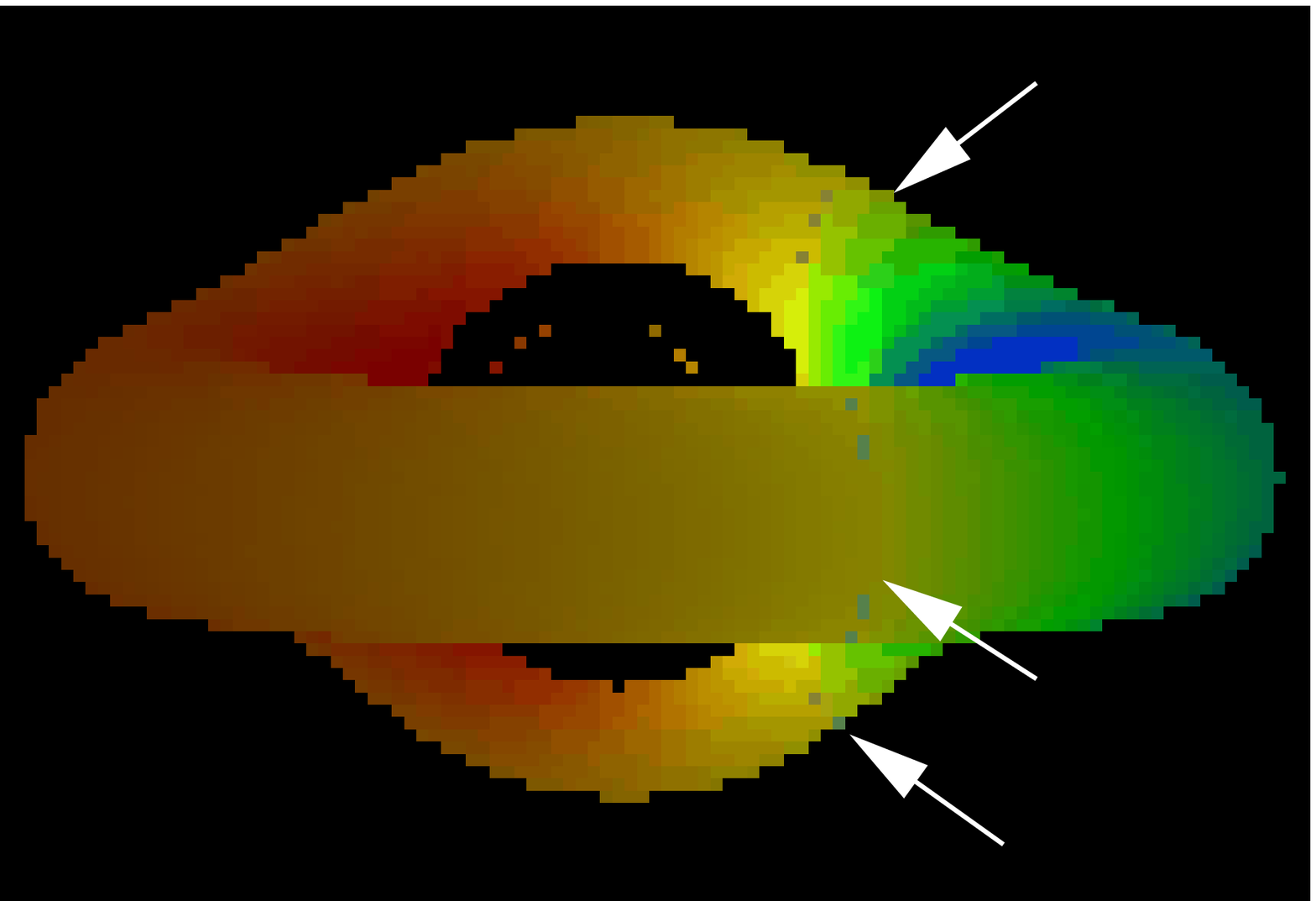, width=7.5cm}
}
\vspace*{0.35cm} 
\caption{
  Images of an accretion torus around a Kerr black hole 
    viewed at inclination angles of $45^\circ$ (top) and $85^\circ$ (bottom).  
  The angular momentum of the black hole $a = 0.998$.    
  The torus is constructed using the velocity law in \S 4.2. 
  The inner and outer radii are determined 
    by the critical surface as described in the same section.  
  High-order images are shown.  
  The energy shift of the emission 
    is represented by a false-colour map laid on the torus surface.   
  Red-shifted emission is coloured yellow/red  
    and blue-shifted emission is coloured blue.     
  \finaldraft{The grey lines indicated by the arrows mark the region  
    where the emission has zero energy shift, 
    separating the torus 
    into regions with red and blue shifted emission.} 
}
\label{torus_image2}
\end{figure}
\subsubsection{Rotational Torus}  

We now investigate the emission from an accretion torus. 
We consider a model in which 
  the inner radius of the torus is determined 
  by the marginally stable orbits of the particles.  
These marginally stable orbits, 
  which depend on $\omega$ and $d \omega/dr$,   
  form a surface in a three-dimensional space. 
The marginally stable orbit for particles in 
  Keplerian motion in the equatorial plane 
  is 6 $R_{\rm g}$  around a Schwarzschild black hole 
  and is 1.23 $R_{\rm g}$ around a Kerr black hole with $a = 0.998$.   
\finaldraft{In equation (\ref{rk-equation}), Keplerian motion corresponds
  to the case with the index $n=0$.
In the torus model that we consider here, $n$ is not
  necessarily zero.
As a consequence, the location of the marginally stable orbit
  of particles in the tori and conventional Keplerian
  disks are different. (See Fig.\ \ref{various_surface}. in \S 4.2).}
We use a surface-finding algorithm to determine the boundary of the torus 
  (see Appendix \ref{surface_finding}).

\begin{figure}
\center{
\epsfig{figure=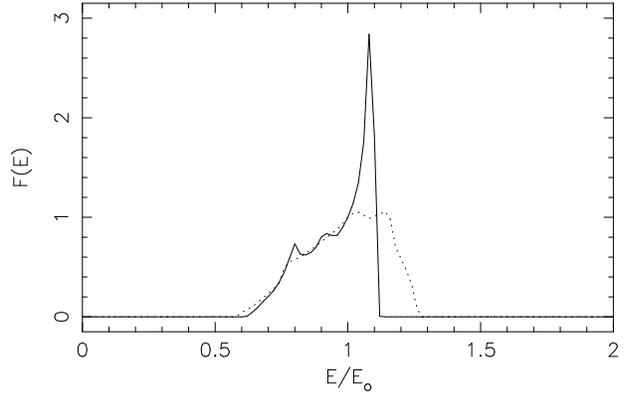, width=8.5cm}
}
\caption{
  Profiles of emission lines 
    from an accretion torus around a Kerr black hole 
    with $a=0.998$ 
    at viewing inclination angles 
    of $45^\circ$ (solid line) and $85^\circ$ (dotted line). 
  The parameters of the torus are $n=0.21$ and $r_k=12R_{\rm g}$ 
    in the velocity profile equation (\ref{rk-equation}).
}
\label{torus_line1}
\end{figure}

\begin{figure}
\center{

\epsfig{figure=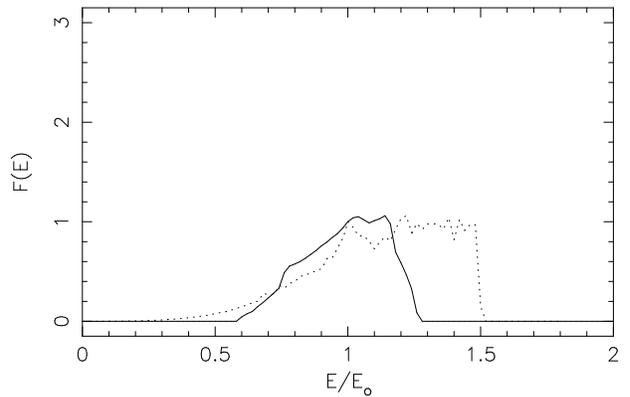, width=8.5cm}
}
\caption{
  A comparison of the profiles of lines 
    from an accretion disk (dotted line)
    and an accretion torus (solid line) 
    around a Kerr black hole with $a = 0.998$. 
  The disk and the torus are  
    viewed at an inclination angle of $85^\circ$. 
  The parameters of the torus are the same as 
     those in Fig. \ref{torus_line1}.  
  The disk has an outer radius of $20R_{\rm g}$,  
    an inner radius of $1.23R_{\rm g}$,
    and a Keplerian velocity profile.
}
\label{torus_line2}
\end{figure} 
 
In Fig. \ref{torus_image2}. we show three-dimensional images 
  of the model torus around a Kerr black hole.  
We include the first four image orders.   
The inclusion of high order images is mandatory 
  due to the `mixing' caused by 
  the extension out of the equatorial plane (Viergutz 1993).    
The torus is viewing inclination angles of 
  $45^\circ$ and $85^\circ$ (top and bottom panels respectively).   
The left-right asymmetry is caused by inertial-frame dragging.  
The multiple images are consequences of gravitational lensing.  
At small inclination angles,  
  only the surface above the equatorial plane of the torus 
  is seen in the direct image. 
At very large inclination angles,  
  the surface below the equatorial plane is severely lensed 
  and also becomes visible.
  
The false-colour map laid on the torus surface 
  show the energy shift of the emitted photons  
  (determined by equation [\ref{freqshift}]),  
  as viewed by a distant observer.   
The separatrix, which corresponds to zero energy shift, 
  divides the torus surface 
  into regions of blue energy shift and regions of red energy shift. 
At large inclination angles 
  the inner surface of the near side of the torus is not visible, 
  and the inner surface of the far side is obscured 
     by the near side of the torus.
Thus, emission with the largest energy shifts is hidden.    
This is very different to the situation 
  for a planar accretion disk 
  -- regardless of the viewing inclination and the visual distortion, 
  the emission from the innermost part of the disk
  is always visible.  
 
Figure \ref{torus_line1}. shows the resulting line profiles 
  obtained by integrating the emission over the images 
  shown in Fig. \ref{torus_image2}.   
The line profile of a torus viewed at $45^\circ$ 
  is similar to that of the planar accretion disk. 
It has a sharp blue peak and smaller red peak. 
It also has an extended red wing.  
This is due to the fact that the projections 
  of a torus and a disk on the sky plane 
  are very similar at low inclination angles.  

However, our calculations show that 
  geometric effects are very important 
  for large viewing inclination angles.   
When some part of the emission region is self-obscured.  
  the resulting profile, as observed from infinity,
  to be completely different from that of the flat disk 
  (see Fig. \ref{torus_line2}.).  
For a torus with large viewing inclination angles, 
  the inner surface of the torus tends to be obscured.  
This corresponds to the region 
  where the most redshifted flux 
  of the line is emitted
 (due to large transverse red shift 
  and gravitational red shift).  
This makes the red wing less prominent. 
The outer surface of the torus is visible from all inclinations.  
As a result the line profile tends to be singly peaked, 
  with the maximum at the unshifted line frequency 
  due to the emission from the outer surface dominating. 
By altering the geometry of the emitter 
  a wide variety of emission profiles can be obtained.

\subsection{Resonant Scattering}

We use a disk model to illustrate the resonant scattering effects.  
We assume the inner edge of the accretion disk is 
  given by the marginally stable orbit.  We use an outer disk radius of $20 R_{\rm g}$ in all the disk simulations.  
This was chosen to accentuate the relativistic effects.  
The emission line profile is assumed to be a delta function.   
We collate the light from the first four image orders of the accretion disk.

\begin{figure*}[ht]
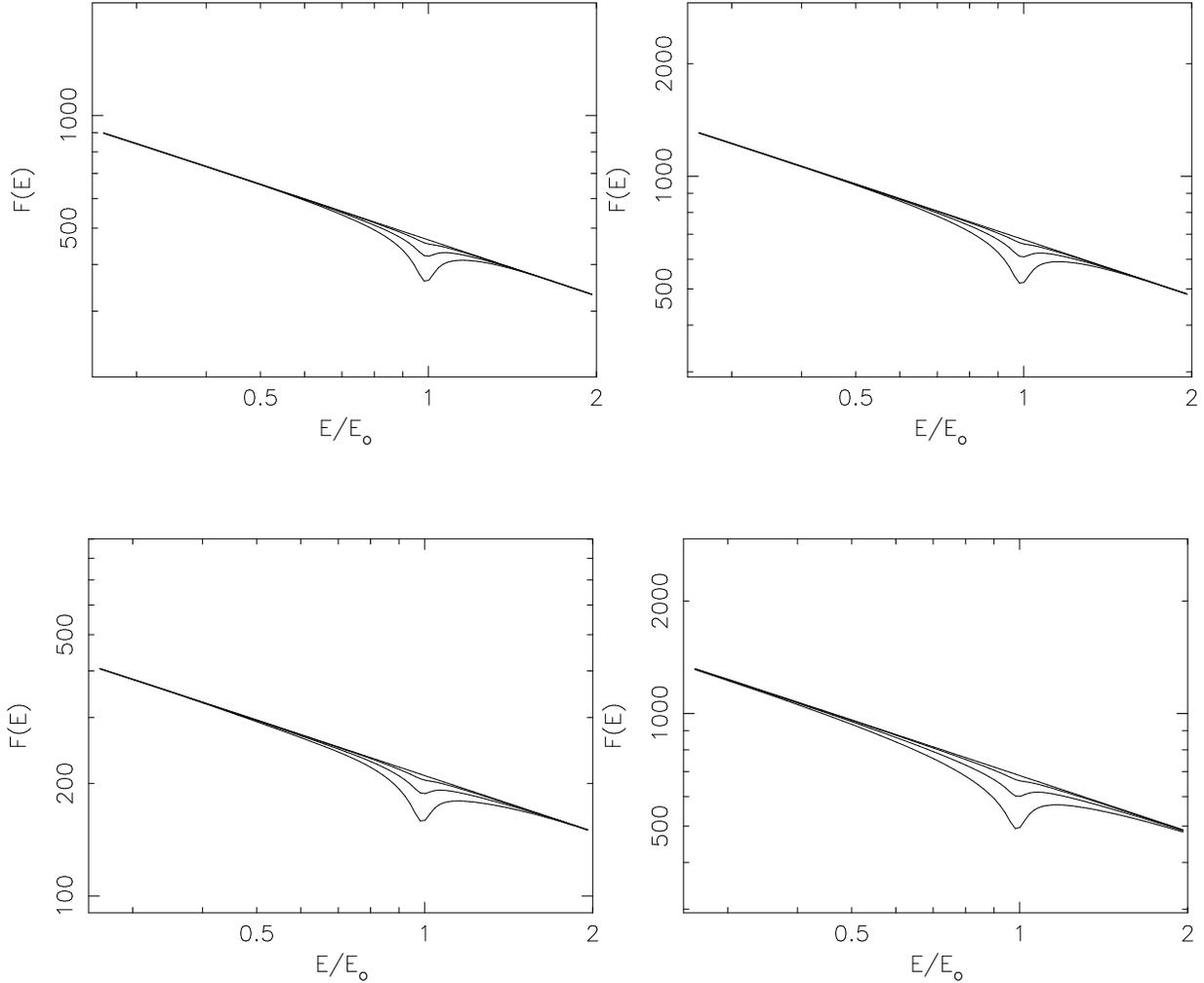

\center{
\epsfig{figure=0814f16.ps, width=8cm}
\epsfig{figure=0814f17.ps, width=8cm}
\epsfig{figure=0814f18.ps, width=8cm}
\epsfig{figure=0814f19.ps, width=8cm}
}
\caption{
Absorbed continuum emission from accretion disks 
   around Schwarzschild and Kerr black holes with $a = 0.998$ 
  (left and right columns respectively), 
  for $i=45^\circ$ and $85^\circ$ (top and bottom panels respectively), 
  $\sigma_0=0$, 0.05, 0.2 and 0.5 
  (curves from top to bottom in each panel).
}
\label{contnoline}
\end{figure*}
\vspace*{0cm}

\begin{figure*}[ht]
\center{
\epsfig{figure=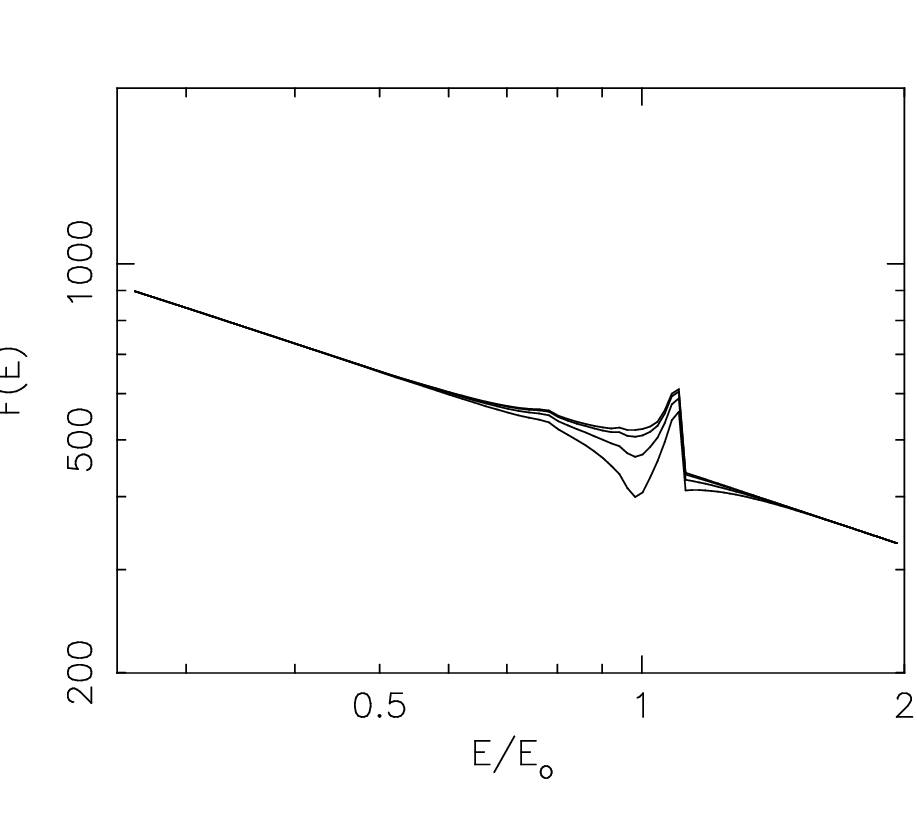, width=8cm}
\epsfig{figure=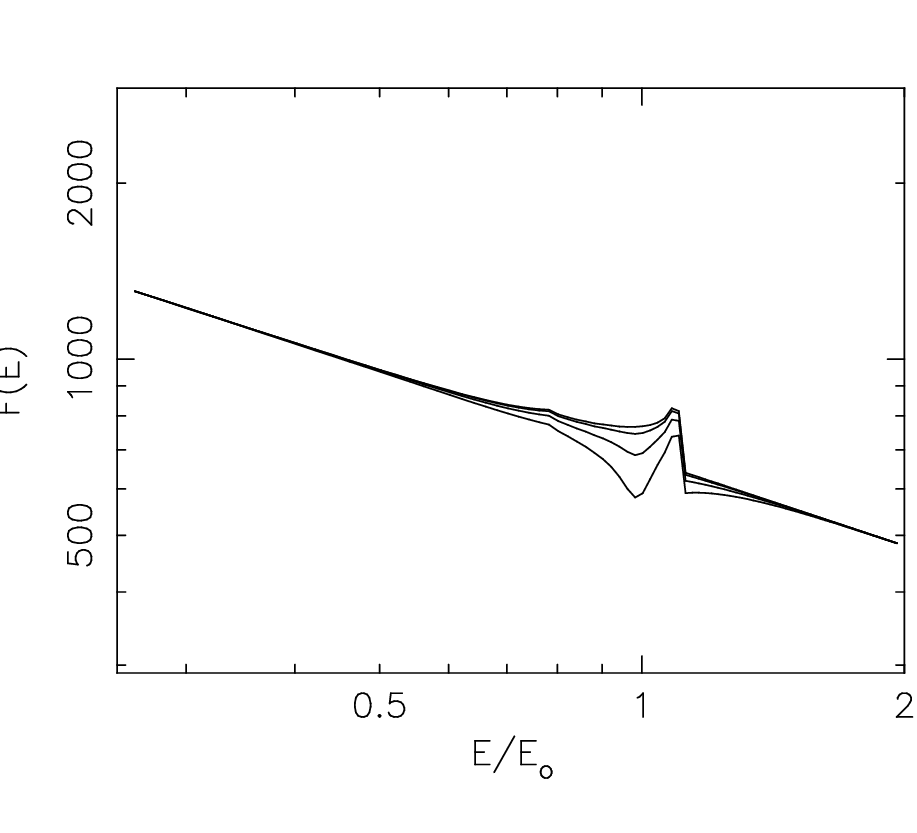, width=8cm}
\epsfig{figure=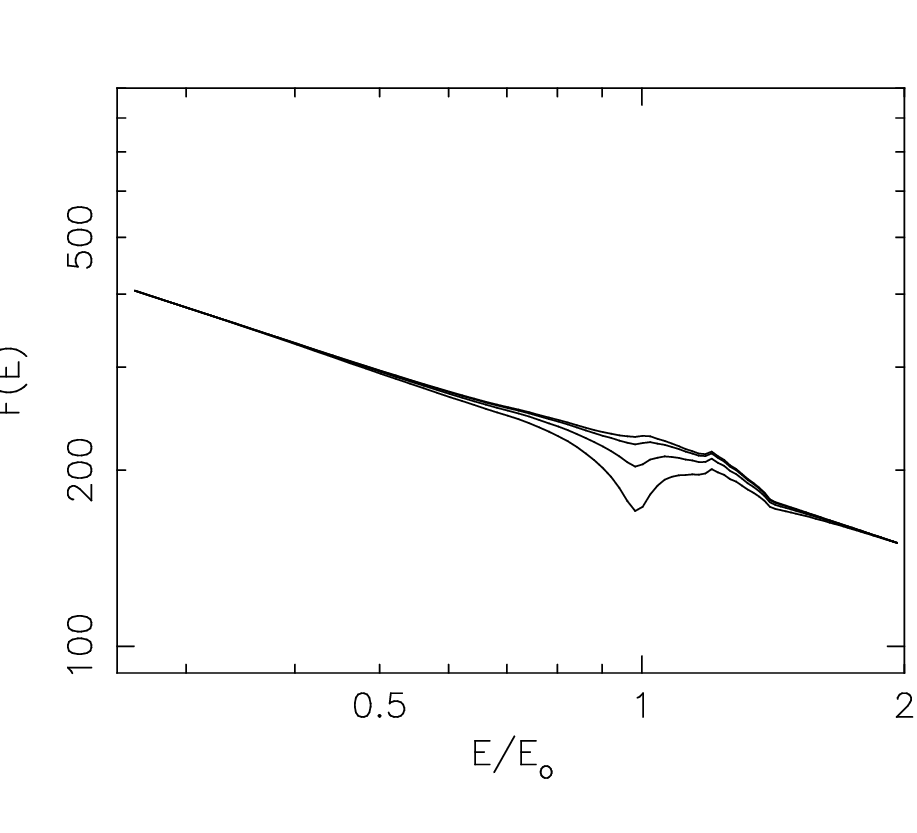, width=8cm}
\epsfig{figure=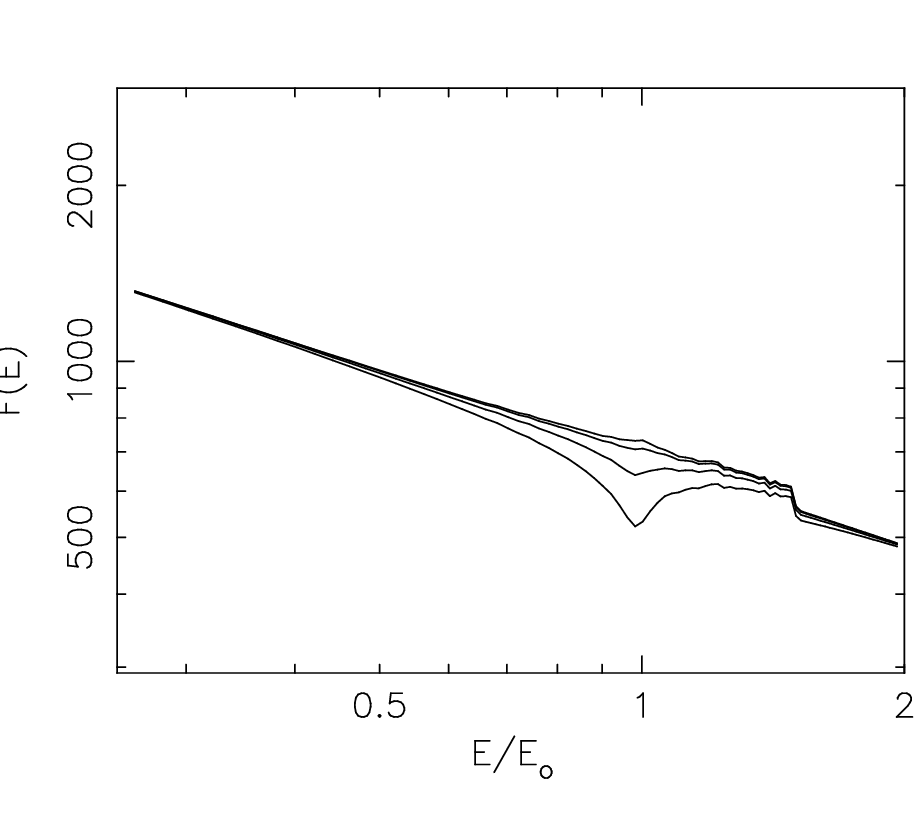, width=8cm}
}
\caption{
Line and continuum emission from accretion disks 
  around Schwarzschild and Kerr black holes with $a = 0.998$ 
  (left and right columns respectively), 
  for $i=45^\circ$ and $85^\circ$ (top and bottom panels respectively), 
  $\sigma_0=0$, 0.05, 0.2 and 0.5 (curves from top to bottom in each panel). 
}
\label{contline}
\end{figure*}
\vspace*{0cm}

The line profiles are obtained from $750\times750$ pixel images.  
The intensity scale on the graphs is in arbitrary units. 
(It is just the log of the sum of pixel intensity over the image, as a function of frequency.)  
The x-axis of the graph is in units of the line rest energy, 
with $E/E_0 = 1$ corresponding to the unshifted line.

We bin the injection spectrum and the absorption coefficient linearly with energy.
There are 1000 bins from $E=0$ to $E=2E_0$.
\finaldraft{Since the thermal width of the line is small,
we assume that the emission line is narrow,
with a width of only one energy bin.
After the calculation, 
we smooth over sets of ten adjacent bins to remove numerical noise
which has a large scatter.
See Appendix \ref{rtalgorithm} for more information about the algorithm used.}

We investigated two space-time models.  
One with $a=0$ corresponding to a Schwarzschild black hole, 
and one with $a=0.998$, corresponding to a maximally spinning Kerr black hole.  
We have plotted spectra containing the continuum 
and the continuum plus line as we have varied the opacity of the absorbing clouds.  

We have also shown how the spectra change with inclination 
due to geometric effects.  We have included a nearly edge-on model 
($i=85^\circ$), and a model with a moderate inclination of $45^\circ$.  See Fig.  \ref{contnoline}. for the results of absorbing the power law continuum, and Fig. \ref{contline}. for the results of absorbing both the line and continuum.

\section{Discussion} 

We have modelled line profiles from accretion disks  
  to demonstrate the use of a general formulation for transfer of radiation  
  through relativistic media in arbitrary space-times. 
In this paper, we used the transfer of emission from AGN as an illustration.  
In this model, we parametrized the disk/torus to describe the emitters 
  and the space-density distribution of the absorbers. 
We took into account relativistic effects 
  on the bulk dynamics and the microscopic kinematic properties 
  of the absorbing medium.  

Resonant absorption/scattering of line emission 
  from accreting black holes in the general relativistic framework 
  had been investigated previously by Ruszkowski \& Fabian (2000).   
In their study, a thin Keplerian disk was assumed 
  and the absorbing medium is a spherical corona of constant density 
  centred on the black hole. 
The corona is rotating, with local rates obtained by linear interpolation 
  from the rotation rate of a planar Keplerian accretion disk 
  and the rotation rate at the polar region 
  caused by frame dragging due to the Kerr black hole.     
The Sobolov approximation was used 
  in the resonant absorption calculations,  
  and a Monte Carlo method determined the re-emission/scattering.  

Our calculation is different to that of Ruszkowski \& Fabian (2000) 
   in the following ways.  
Firstly, the emitters are not confined to the equatorial plane, 
  i.e. they can be thin accretion disks or thick tori.   
Secondly, the absorbing medium is a collection of (cold) clouds 
  with relativistic motions.  
The number density distribution of the clouds 
  is parametrized by a powerlaw decreasing radially.
The local bulk (rotational) velocity  of the clouds   
  is determined by general relativistic dynamics, 
  and the velocity dispersion is calculated from the Virial theorem.    
Thirdly, we do not assume the Sobolov approximation.  
The resonant condition for the absorption coefficient is derived directly 
  from the kinematics of the absorbing cloud particles.  
Fourthly, we ignore the contribution from re-emission to the line flux.  
However, we include emission 
  from higher order disk images, in addition to the direct image.   
 
One of the main differences between the two studies  
  is the treatment of resonant absorption.  
In the Sobolov approximation,  
  the absorption takes place locally    
  (see Rybicki and Lightman 1979). 
The line profile is practically a delta function; 
  otherwise, the assumption of quasi-local absorption breaks down.    
Moreover, it requires that the absorbing medium is a radial flow. 
The emission lines are \finaldraft{broadened} because of relativistic effects, 
  and the motion of the medium is rotationally dominated. 
The locality of the absorption (required by the Sobolov approximation) 
  therefore breaks down.   
To overcome these difficulties, 
  we abandon the Sobolov approximation 
  but, instead, employ full ray tracing.  
    
In modelling the bulk flow, 
  we derived the equation of motion of the absorbers 
  using the rotational-support approximation for the accretion disk. 
In addition, we assume that the flow is supported out of the equatorial plane.  
There is negligible radial force in the equations of motion,
  and the radial velocity can be neglected. 
To include outflowing as a wind, or inflowing, 
  requires additional components in the equation of motion.  
This complicates the formulation, in particular, 
  when matching the flow boundary condition at the surfaces 
  of the accretion disk/torus. 
A self-consistent boundary condition requires a dissipation mechanism 
  in the boundary regions of the disk/torus and absorbing clouds.   
The inclusion of such dissipation is beyond the scope of this paper 
  and this issue will be addressed in future works.  

As the rays propagate from the accretion disk to the observer, 
  they experience position-dependent absorption.  
The absorption depends upon the velocity profile of the material 
  as well as its density and the line profile function 
  of the absorption coefficient.  
Since the absorbers are moving in relativistic speeds, 
  the bulk velocity is an important factor.  
Lorentz contraction increases the absorption coefficient accordingly, 
  and Doppler shift alters the frequency of the emission 
  as seen by the absorbers.   

The potential energy liberated by material infalling into a black hole 
  is of order a few percent of the material's rest mass,  
  and the energy corresponding to `thermal' kinematic velocity dispersion 
  approaches this rest mass energy. 
Thus, we replace the conventional Maxwellian distribution 
  by the J\"{u}ttner distribution for relativistic particles 
  in deriving the resonant line absorption coefficient.   
This distribution does not give a Gaussian absorption profile 
  even in the bulk rest frame of the absorbers.  

The \finaldraft{absorption} line profiles of the direct images 
  show a dip around the line rest frequency (energy) $(E/E_0 =1)$. 
Absorption can change the line profile significantly.  
The flux around the rest frequencies  
  is, however, augmented by the flux from shifted lines 
  from the higher order images, 
  especially at high disk inclination.  
These two effects compete, 
  and when the line-of-sight optical depth is high, 
  the contribution of the high-order images is masked.    

\section{Conclusion}

We present a numerical ray-tracing method 
  for radiative-transfer calculations in curved space time 
  and apply the method to calculate line emission 
  from accretion disks and tori around black holes.  
Our calculations have shown 
  that lines from relativistic accretion tori  
  have profiles very different 
  to lines from relativistic thin planar accretion disks 
  for the same system parameters, 
  such as the spin of the black hole and the viewing inclination. 
The self-obscuration of the inner region of the accretion torus  
  leads to weaker red wing in the emission lines 
  when compared with the lines emitted from a thin planar accretion disk.  
At high inclination angles 
  the strong blue peak is also absent in the line emission from the tori.   
   
We also investigate the effects of resonant absorption/scattering 
  by the line-of-sight material in relativistic motion 
  with respect to the emitters in the disk/torus, and the observer.  
Our method does not invoke the Sobolov approximation, 
  and the resonant absorption/scattering condition is derived directly.
We have shown that absorption effects are important 
  in shaping the profiles of emission lines. 
The interpretation of observations of relativistic lines from AGN 
  is non-trivial when absorption is present.

\begin{acknowledgements}

We thank Mat Page for discussion,
  and Alex Blustin for comments on the manuscript.
SVF acknowledges the support 
  by a UK government Overseas Research Students Award
  and a UCL Graduate School Scholarship.
\finaldraft{We thank the referee for clarifying the issue regarding
  energy redistribution in the scattering process.}

\end{acknowledgements}

\appendix    

\section{Frequency Bins}
Using an astrophysical model together with the metric, we can obtain an equation describing the flow of the medium through which the ray is propagating; $u^\alpha(x^\alpha(\lambda))$.  The properties and distribution of the medium can be used to derive the absorption and emission terms $\chi_0(x^\alpha(\lambda))$ and $\eta_0(x^\alpha(\lambda))$.  These can then be used to calculate what is observed from infinity via equation (\ref{anaray}).  In practice this simple method is altered with a ray-tracing numerical algorithm.  The radiative transfer equation is not time-symmetric, so if one follows rays back from the observer to the emitter the above formulation cannot be used.  The emission frequency and intensity is not known until the emitter is reached along the path.

This problem can be solved using one of two approaches.  The first is to collate the path from emitter to observer, and then to integrate along that path forward in time using equation (\ref{anaray}).  The second approach is to collate the optical depth for $\cal I$ for a table of frequencies as one travels back in time.  This optical depth can then be used to calculate the observed intensity, $I$, for each binned frequency.  The results for each emission element along the path are simply summed.

Since, in general, the resulting frequencies are binned anyway to get a spectrum, we choose the second method.  It requires less storage, and is a simple extension of the grey-absorber case where $\chi_0$ and $\eta_0$ are no longer functions of $\nu$.  In our treatment, we sum over the observed frequencies rather than those in the rest frame, avoiding the need to transform the distribution of the frequencies along the path due to the gravitational redshift factor.  The act of going from the emission frame to the observer's frame also affects the intensity $I$, so we use the invariant $\cal I$ instead, only converting to specific intensity just before outputting the results.

Binning the optical depth with frequency along a path also allows us to model the radiative transport of the continuum as well as the lines, provided that one is in the limit where scattering is unimportant.  (Stimulated emission can be modeled by using a negative absorption coefficient.)  Depending on the relative intensities of the continuum and the line, absorption may cause the emission feature to be converted into an absorption feature.

We assume that the continuum is a power law, which we parametrize by a slope and intensity at the line rest frequency.  We have assumed that the continuum is emitted with an intensity that scales with the line emissivity.  This means that the equivalent width of the emission line is constant across the disk.  We set this to be $0.05$ of the rest line frequency.  (This value is roughly what is seen in AGN.)   To simplify things further, we will fix the power law index of the continuum to be $\gamma=0.5$, where $\gamma$ is defined by
\begin{equation}
I=C E^{-\gamma},
\end{equation}
in which $I$ is the continuum intensity as a function of energy, $E$, and $C$ is derived from the given equivalent width.  This rather hard spectrum was chosen to emphasise the line.

The treatment assumes that the continuum is created in a relatively thin planar structure above the accretion disk.  In effect, we treat the emission from the disk corona as part of the injection spectrum, together with the emission line, which we propagate through the absorbing material suspended much higher above.

\section{Ray Tracing Algorithm}
\label{rtalgorithm}
A direct ray-tracing method is used instead of the conventional transfer-function method, as it is easier to incorporate the numerical radiative-transport calculations.  The ray-tracing algorithm is as follows:

\begin{enumerate}
\item Integrate the equations governing the geodesics, and those describing the optical depth for each frequency, from the observer to the emitting surface;
\item At each crossing of the equatorial plane / torus surface, collate the position and the direction of the photon; 
\item Construct the image, and determine the observed frequency/energy shift;
\item Integrate the emission over the images of each order to produce the line profiles.
\end{enumerate}

\finaldraft{We use a Runge-Kutta integrator to calculate the ray trajectories.  The relative tolerance is set to $10^{-11}$, which prevents the ray tracer from missing the torus by taking steps which are too large.  It also allows a simple Eulerian method to be used to integrate the optical depths.  The absorption coefficient is a slowly varying function on the scale of the step size, which allows this optimisation.
}

The foot points of the null geodesics (photon trajectories) on the disk surface are calculated by a root-finding algorithm  (see e.g.\ Press et al.\ 1992, p.343). The first four intersections of the null geodesics and the disk plane (corresponding to the direct and first three higher-order images), and the four-vectors of the photons emitted from there are recorded.  The incorporation  of an emissivity law is therefore straightforward, as it is defined in terms of the spatial coordinates on the disk plane.  Since the trajectory of the photon is also saved at each crossing point, it is also possible to include the effects of limb darkening, and to model a semi-transparent disk.

Since the disk is imaged upon a sky plane, all gravitational lensing effects on the intensity of the light are implicitly included in the calculation of the image itself.  If a region of the disk is magnified then it will cover more area in the image, and thus will appear brighter than a non-magnified region.  Inclination effects are also included implicitly.  An inclined disk will cover less pixels than a face-on disk, where the number of pixels is roughly proportional to $\cos i$, where $i$ is the inclination angle.  (Light bending causes this Euclidean formula to be only an approximation.)  Gravitational lensing does not alter the observed surface brightness of a point.

This implicit inclusion of changing areas of photon flux-tubes linking the observer to the emitter vastly simplifies the calculation of the observed flux.  All that is required is to integrate over each pixel on the image, taking into account the redshift of the emission regions corresponding to the pixels.  If one were integrating over the surface of the disk, instead of over the image, then the Jacobian of the transformation from the disk to the image plane coordinates would be required.  This is numerically difficult to obtain, and would require a separate transformation for each image order.

\finaldraft{The programs (written in Fortran) take a few hundred kilobytes of memory under Tru64 Unix on a OSF1 V5.1 1885 Alpha.   One version outputs the intensity / frequency shift of the emission line for each `pixel' in the sky plane.  Another version collates the continuum as well as the line, and outputs a table of intensity verses energy integrated over the whole disk.  Depending on load on the shared system, and on the image size requested, a typical run takes a few hours per image.}

\section{Surface Finding Algorithm} 
\label{surface_finding} 

We consider the following algorithm to determine the torus surface.  
We integrate equation (\ref{isobaric}) 
  and tabulate the resulting points ($r,\theta$) 
  along the path of the integration.  
Then we interpolate ($r,\theta$) 
  and construct the torus surface, 
  where the emission originates.  
We use spline interpolation between the surface points. 

When ray tracing the photon paths,
  we determine the intersection  of the trajectory 
  and the torus surface.
As the photon trajectory calculations may take large spatial steps, 
  there is a possibity that the torus is not 'detected'. 
To prevent this from happening,   
  we consider the following procedure.
We take note of the region where the photon is located 
  during the trajectory calculation: 
  either inside or outside the torus 
  and either above or below the equatoral plane. 
Whenever the photon leaves one of the four regions,
  and enters another region, 
  we use a boundary-searching algorithm  
  to find the exact location where the transit occurs.  
If the trajectory hits the equatorial plane outside the torus, 
  the integration will continue. 
If the trajectory hits the torus, then integration is terminated. 

By taking smaller steps 
  required by the boundary-finding algorithm, 
  we can prevent the integrator from missing the torus entirely.  
This algorithm can be used for more complicated surfaces.
It works well, because the integrator will only miss intersections
  when the trajectory is close to tagential to a surface.
This happens close to the equatorial plane in the torus models.
Adding in a fake boundary there,
  and thus decreasing the step size,
  helps in preventing missed intersections.
\end{document}